\documentclass[aps,prb,twocolumn,superscriptaddress,showpacs]{revtex4-1}

\usepackage[dvipdfmx]{graphicx}
\usepackage{graphicx}
\usepackage{amsmath}
\usepackage{amssymb}
\usepackage{bm}
\usepackage{color}
\usepackage{here}
\usepackage{comment}
\usepackage{multirow}

\begin{document}

\title{
Theory of light-induced resonances with collective Higgs and Leggett modes \\ in multiband superconductors
}

\date{\today}
\author{Yuta Murotani}
\affiliation{Department of Physics, University of Tokyo, Hongo, Tokyo 113-0033, Japan}
\author{Naoto Tsuji}
\affiliation{RIKEN Center for Emergent Matter Science (CEMS), Wako 351-0198, Japan}
\author{Hideo Aoki}
\altaffiliation[Presently also at ]{Electronics and Photonics Research Institute,
Advanced Industrial Science and Technology (AIST),
Tsukuba, Ibaraki 305-8568, Japan.}
\affiliation{Department of Physics, University of Tokyo, Hongo, Tokyo 113-0033, Japan}
\affiliation{High Energy Accelerator Research Organization (KEK), Tsukuba,
Ibaraki 305-0801, Japan}

\pacs{74.40.Gh, 74.25.N-, 74.25.Gz, 74.70.-b}

\begin{abstract}
We theoretically investigate coherent optical excitations 
of collective modes in two-band BCS superconductors, which accommodate two Higgs modes and one Leggett mode corresponding, respectively, to the amplitude and relative-phase oscillations of the superconducting order parameters associated with the two bands.  We find, 
based on a mean-field analysis, that each collective mode can be resonantly excited through a nonlinear light-matter coupling when the doubled frequency of the driving field coincides with the frequency of the corresponding mode.
Among the two Higgs modes, the higher-energy one exhibits a sharp resonance with light, 
while the lower-energy mode has a broadened resonance width.  
The Leggett mode is found to be resonantly induced by a homogeneous ac electric field 
because the leading nonlinear effect generates a potential offset between the two bands that couples to the relative phase of the order parameters.
The resonance for the Leggett mode becomes sharper with increasing temperature.  
All of these light-induced collective modes 
along with density fluctuations contribute to
the third-harmonic generation.  
We also predict an experimental possibility of
optical detection of the Leggett mode.
\end{abstract}

\maketitle

\section{Introduction}
Since collective modes go hand in hand with spontaneous symmetry breaking, they are one of the best probes of many-body systems.  
When a continuous symmetry is spontaneously broken, a massless Nambu-Goldstone (NG) mode\cite{Nambu,Goldstone1,Goldstone2} should appear in general.
In the case of $U(1)$ symmetry breaking such as neutral superfluid $^3$He and superconductors, it takes the form of an excitation of the phase of the order parameter.  
In superconductors, however, electrons, being charged, 
are coupled to the electromagnetic field, so that the NG mode is elevated to a high energy due to the Anderson-Higgs (AH) mechanism\cite{Anderson,Higgs,Englert,Higgs2,Gralnik}, making it difficult to be observed.  
In the vicinity of the superconducting phase transition, the massless NG mode energy can be low when the superfluid and normal components coexist and cooperatively propagate in the form of Carlson-Goldman mode\cite{Carlson,Ohashi}. In addition to these, fluctuations in the amplitude of the order parameter exist\cite{Volkov} as well, and their collective excitation is called Higgs mode\cite{Higgs2,Varma,Pekker} when the system is coupled to gauge fields.  Existence of the Higgs mode in a conventional superconductor has been confirmed with Raman spectroscopy\cite{Sooryakumar,Littlewood,Measson}, and more recently with terahertz (THz) spectroscopy\cite{Matsunaga1,Matsunaga2,Tsuji}.

Now, if we go over to multi-component superconductors, where the superconducting order parameter consists of multiple complex components, 
we can expect 
they should accommodate versatile collective modes.  
Indeed, superfluid $^3$He is known to have multiple amplitude modes coming from spin-triplet and $p$-wave nature of Cooper pairs\cite{Wolfle}. 
For $d$-wave superconductors such as high-$T_c$ cuprates, it has been group-theoretically shown\cite{Barlas} that they can 
accommodate additional amplitude (Higgs) modes coming from 
multiple irreducible representations of the $k$-dependent 
gap function with the $D_4$ point-group symmetry.  
As for the phase modes, multi-gap superconductors are predicted to 
have an out-of-phase mode between the 
two components of the gap function\cite{Leggett,Sharapov,Burnell,Ota,Lin,Marciani,Bittner,Krull,Cea2}, called ``Leggett mode''.

MgB$_2$ is a typical example of multi-gap superconductors\cite{Szabo,Iavarone}.
Its double-gap structure
originates from an electronic structure around the Fermi energy that 
comprises $\sigma$ and $\pi$ bands\cite{Kortus,Liu,Souma}.
Observation of the Leggett mode in MgB$_2$ has been reported with tunneling spectroscopy\cite{Brinkman}, Raman spectroscopy\cite{Blumberg,Klein} and angle-resolved photoemission spectroscopy\cite{Mou}, while no report has so far been made for the Higgs mode.   
A more recent family of superconductors, the iron pnictides with 
high $T_c$s, 
also have multi-orbital and multi-gap structures, 
where the electron correlation is suggested\cite{Kuroki,mazin} 
to bring about $s_{\pm}$ and $s_{++}$ pairings depending on the chemical composition 
and/or doping level\cite{Kuroki2,shibauchi}. 
Study of collective modes in such multi-band superconductors should shed a new light on their order parameter and pairing interactions. 
For instance, it has been predicted that competing $s$- and $d$-wave interactions can result in different collective modes for different ground states\cite{Marciani,Maiti}.
Collective modes are also recently studied 
for systems where superconductivity coexists with diagonal orders 
such as spin-density wave\cite{Moor1,Dzero} or charge-density wave\cite{Moor2}.

Multi-component superconductors thus accommodate 
a variety of collective modes, but we are still in need of 
a systematic study for them,
where the Leggett and Higgs modes should be 
simultaneously examined by varying relative sizes and the 
coupling of multiple 
superconducting gaps.  This has motivated us to specifically pose a question: 
how are the Higgs and Leggett modes coupled to electromagnetic fields in multi-gap superconductors?  
In the single-band case, the Higgs mode couples to gauge fields nonlinearly\cite{Tsuji}, 
which makes it possible to optically excite the mode, typically with an intense THz laser\cite{Matsunaga2}. Now, 
for multi-component superconductors we shall reveal, 
based on a mean-field analysis, 
that each collective mode can be resonantly excited through a nonlinear light-matter coupling when the doubled frequency of the driving field coincides with the frequency of each mode, as in the single-band case.  
More importantly, we shall show that each of the two Higgs modes and the Leggett mode exhibits
dramatically different sharpness in the resonance, depending on
the interband pairing interaction 
(i.e., interband Josephson coupling) and temperature.
The resonance itself can be interpreted as two-photon absorption by collective modes, 
which contrasts with Raman scattering where the energy difference between the incident and scattered photons is absorbed by elementary excitations including collective modes.

Second purpose of the present work is to 
examine how the light-induced Higgs and Leggett modes contribute to nonlinear optical responses, especially to the third-harmonic generation (THG).
The resonantly induced THG at the frequency of half the superconducting gap has been observed experimentally for NbN, a single-gap superconductor\cite{Matsunaga2}.
The THG resonance is contributed from the Higgs mode\cite{Tsuji} and density fluctuations, the latter being pointed out to be dominant within the BCS mean-field theory\cite{Cea}.  
In this paper, we examine THG arising from the Higgs modes and density fluctuations in two-band superconductors. They 
are shown to have distinct resonance features, which will open a way 
to experimentally probe multi-band superconductors.
We also point out another THG feature specific to multi-band cases arising from the Leggett mode, where we shall discuss the possibility of detecting the Leggett mode through THG measurement.

This paper is organized as follows. 
In Sec. II we construct a dynamical theory for two-band superconductivity in the BCS regime. 
In Sec. III we calculate the response of the relative phase to an ac electric field to derive the optical resonance of the Leggett mode. Section IV is devoted to the Higgs amplitude modes and their optical resonances. 
Section V examines the effects of finite temperatures on these modes. In Sec. VI we 
reveal how the light-induced collective modes will appear in THG.
We summarize the results and future prospects in Sec. VII.

\section{Pseudospin representation for two-band superconductors} \label{section2}

Let us first derive the equation of motion for optically excited two-band superconductors, in terms of Anderson's pseudospins. We start with the Hamiltonian,
\begin{align}
&\mathcal{H}=\sum_{\bm{k}\sigma}\epsilon_{\alpha(\bm{k}-e\bm{A}(t))}\alpha^\dag_{\bm{k}\sigma}\alpha_{\bm{k}\sigma}
+\sum_{\bm{k}\sigma}\epsilon_{\beta(\bm{k}-e\bm{A}(t))}\beta^\dag_{\bm{k}\sigma}\beta_{\bm{k}\sigma} \notag \\
&+ \sum_{\bm{kk}'}\left[V_{\alpha\alpha}\alpha^\dag_{\bm{k}\uparrow}
\alpha^\dag_{-\bm{k}\downarrow}\alpha_{-\bm{k}'\downarrow}\alpha_{\bm{k}'
\uparrow}
+V_{\beta\beta}\beta^\dag_{\bm{k}\uparrow}
\beta^\dag_{-\bm{k}\downarrow}\beta_{-\bm{k}'\downarrow}\beta_{\bm{k}'
\uparrow}\right. \notag \\
&\left.
+\left(V_{\alpha\beta}\beta^\dag_{\bm{k}\uparrow}
\beta^\dag_{-\bm{k}\downarrow}\alpha_{-\bm{k}'\downarrow}\alpha_{\bm{k}'\uparrow}
+{\rm h.c.}\right)\right],
\label{H_2band}
\end{align}
where subscripts $\alpha$ and $\beta$ label the two bands, 
$\alpha^\dag_{\bm{k}\sigma} (\beta^\dag_{\bm{k}\sigma})$ creates an electron 
with momentum $\bm{k}$ and spin $\sigma$ in band $\alpha (\beta)$, 
$\epsilon_{\alpha\bm{k}}$ and $\epsilon_{\beta\bm{k}}$ are respective 
band dispersions measured from the chemical potential,
$V_{\alpha\alpha}$ and $V_{\beta\beta}$ are respective intraband pairing interactions,
while $V_{\alpha\beta} (=V^\ast_{\beta\alpha})$ is the interband 
pairing interaction. $\bm{A}(t)$ is the vector potential 
representing the laser field, which is assumed to be spatially 
homogeneous, i.e., the superconductor is assumed to be 
thinner than the penetration depth and the wavelength of light.  
Optical interband transitions are neglected here, because we consider the incident light (such as THz waves) with energies much lower than the interband 
transitions.  
We further ignore differences in the microscopic charge distribution of Wannier orbitals between $\alpha$ and $\beta$ bands. 
In this approximation, it is known that the Leggett mode at zero momentum is not affected by the 
Anderson-Higgs (AH) mechanism\cite{Leggett, Sharapov, Burnell, Bittner}, since the interband charge transfer associated with the Leggett mode does not induce an electric current in real space. Hence the Leggett mode is not coupled linearly to electromagnetic fields, and survives at low energies.  
Even when the difference in the orbital charge distributions is taken into account, 
it will not contribute to the long-wavelength screening (i.e., the AH mechanism), since the interband current will only 
occur over typical wave vectors associated with the size of Wannier orbitals.
Therefore we adopt the Hamiltonian (\ref{H_2band}) in the present paper.

Let us then define the mean fields,
\begin{equation}
\Psi_{\alpha\bm{k}} \equiv \langle\alpha^\dag_{\bm{k}\uparrow}\alpha^\dag_{-\bm{k}\downarrow}\rangle,~
\Psi_{\beta\bm{k}} \equiv \langle\beta^\dag_{\bm{k}\uparrow}\beta^\dag_{-\bm{k}\downarrow}\rangle,
\label{mean_field}
\end{equation}
and
\begin{align}
\Delta_\alpha=-V_{\alpha\alpha}\sum_{\bm{k}}\Psi_{\alpha\bm{k}}-V_{\alpha\beta}\sum_{\bm{k}}\Psi_{\beta\bm{k}},
\nonumber \\
\Delta_\beta=-V_{\beta\alpha}\sum_{\bm{k}}\Psi_{\alpha\bm{k}}-V_{\beta\beta}\sum_{\bm{k}}\Psi_{\beta\bm{k}},
\label{gap}
\end{align}
which yield a two-band BCS Hamiltonian,
\begin{equation}
\mathcal{H}_{\mathrm{BCS}}=\mathcal{H}_\alpha+\mathcal{H}_\beta, \label{Ham_sum}
\end{equation}
with
\begin{align}
\mathcal{H}_{\gamma}
&=\sum_{\bm{k}\sigma}\epsilon_{\gamma(\bm{k}-e\bm{A}(t))}\gamma^\dag_{\bm{k}\sigma}\gamma_{\bm{k}\sigma}
-\Delta^*_\gamma\sum_{\bm{k}}\gamma^\dag_{\bm{k}\uparrow}\gamma^\dag_{-\bm{k}\downarrow}
\notag
\\
&\quad
-\Delta_\gamma\sum_{\bm{k}}\gamma_{-\bm{k}\downarrow}\gamma_{\bm{k}\uparrow}
+\Delta^*_\gamma\sum_{\bm{k}}\Psi_{\gamma\bm{k}}
\end{align}
for $\gamma=\alpha,\beta$.
Now it is convenient to introduce Anderson's pseudospin\cite{Anderson2},
\begin{equation}
\bm{\sigma}_{\gamma\bm{k}}=\frac{1}{2}(
\begin{array}{cc}
\gamma^\dag_{\bm{k}\uparrow} & \gamma_{-\bm{k}\downarrow}
\end{array})\bm{\tau}\left(
\begin{array}{c}
\gamma_{\bm{k}\uparrow} \\ \gamma^\dag_{-\bm{k}\downarrow}
\end{array}\right), \label{pseudospin}
\end{equation}
where $\bm{\tau}=(\tau^x,\tau^y,\tau^z)$ are the Pauli matrices 
with respect to the Nambu spinors. 
The Hamiltonian is then concisely expressed, up to a constant, as
\begin{equation}
\mathcal{H}_{\mathrm{BCS}}=\sum_{\gamma=\alpha,\beta}\sum_{\bm{k}}2\bm{b}_{\gamma\bm{k}}\cdot\bm{\sigma}_{\gamma\bm{k}},
\end{equation}
where
\begin{equation}
\bm{b}_{\gamma\bm{k}}=\left(-\Delta'_\gamma,-\Delta''_\gamma,\frac{\epsilon_{\gamma(\bm{k}-e\bm{A}(t))}+\epsilon_{\gamma(\bm{k}+e\bm{A}(t))}}{2}\right) \label{pseudomagnetic}
\end{equation}
is a pseudomagnetic field acting on the pseudospins,
with $\Delta'_\gamma$ and $\Delta''_\gamma$ respectively denoting the real 
and imaginary parts of $\Delta_\gamma$. 
The equation of motion for pseudospins then takes a form of the Bloch equation,
$\partial\bm{\sigma}_{\gamma\bm{k}}/\partial t=i[\mathcal{H}_{\mathrm{BCS}},\bm{\sigma}_{\gamma\bm{k}}]=2\bm{b}_{\gamma\bm{k}}\times\bm{\sigma}_{\gamma\bm{k}}$, 
or, for the mean fields, 
\begin{equation}
\frac{\partial\langle\bm{\sigma}_{\gamma\bm{k}}\rangle}{\partial t} = 2\bm{b}_{\gamma\bm{k}}\times\langle\bm{\sigma}_{\gamma\bm{k}}\rangle. 
\label{BdG_eqn}
\end{equation}
We have to solve this equation to self-consistently satisfy Eq.(\ref{gap}), i.e.,
\begin{align}
&\Delta'_\gamma=-V_{\gamma\alpha}\sum_{\bm{k}}\langle\sigma^x_{\alpha\bm{k}}\rangle-V_{\gamma\beta}\sum_{\bm{k}}\langle\sigma^x_{\beta\bm{k}}\rangle,~
\nonumber \\
&\Delta''_\gamma=-V_{\gamma\alpha}\sum_{\bm{k}}\langle\sigma^y_{\alpha\bm{k}}\rangle-V_{\gamma\beta}\sum_{\bm{k}}\langle\sigma^y_{\beta\bm{k}}\rangle, \label{self_consistency}
\end{align}
when $V_{\alpha\beta}$ is real.
We shall suppress brackets denoting the expectation values hereafter.

As the initial state, we take the thermal equilibrium state, which 
can be obtained by diagonalizing the mean-field Hamiltonian (\ref{Ham_sum}) with $\bm{A}=0$ 
and assuming the thermal (Fermi) distribution of the resulting quasi-particles.
In terms of the pseudospins, the thermal state is given by
\begin{align}
\sigma^{x,\mathrm{eq}}_{\gamma\bm{k}}&=\frac{\Delta^{\mathrm{eq}}_\gamma}{2E_{\gamma\bm{k}}}\tanh\left(\frac{E_{\gamma\bm{k}}}{2k_BT}\right),~
\sigma^{y,\mathrm{eq}}_{\gamma\bm{k}}=0, \nonumber\\
\sigma^{z,\mathrm{eq}}_{\gamma\bm{k}}&=-\frac{\epsilon_{\gamma\bm{k}}}{2E_{\gamma\bm{k}}}\tanh\left(\frac{E_{\gamma\bm{k}}}{2k_BT}\right), \label{gap_eqn}
\end{align}
where the superscript ``eq'' denotes the value in equilibrium, $\Delta^{\mathrm{eq}}_\gamma$ chosen real, 
\begin{equation}
E_{\gamma\bm{k}}=\sqrt{\epsilon_{\gamma\bm{k}}^2+(\Delta^{\mathrm{eq}}_\gamma)^2}, \label{QPE}
\end{equation} 
$k_B$ the Boltzmann constant, and $T$ the temperature. 
Again, Eq.(\ref{gap_eqn}) is subject to the self-consistency condition (\ref{self_consistency}). 
This gives the gap equation for two-band superconductors, 
from which we can show that the sign of $V_{\alpha\beta}$ determines the relative phase of the two gaps:
a repulsive interaction $V_{\alpha\beta}>0$ favors $s_\pm$ paring
(defined as those with sign reversal, $\Delta_\alpha\Delta_\beta<0$),
while an attractive $V_{\alpha\beta}<0$ favors $s_{++}$ pairing (with $\Delta_\alpha\Delta_\beta>0$).
Here the terminology of $s_{\pm}$ and $s_{++}$ are adopted
from those in iron pnictide superconductors\cite{Kuroki, Kuroki2}.

When the system is irradiated by a laser with small intensity, we can linearize Eq.(\ref{BdG_eqn}) with respect to the deviations from the equilibrium, 
\begin{align}
\label{x_eqn}
\partial_t\delta\sigma^x_{\gamma\bm{k}}(t)&=-2\sigma^{z,\mathrm{eq}}_{\gamma\bm{k}}\delta\Delta''_\gamma(t)-2\epsilon_{\gamma\bm{k}}\delta\sigma^y_{\gamma\bm{k}}(t), \\
\label{y_eqn}
\partial_t\delta\sigma^y_{\gamma\bm{k}}(t)&=2\epsilon_{\gamma\bm{k}}\delta\sigma^x_{\gamma\bm{k}}(t)+2\sigma^{x,\mathrm{eq}}_{\gamma\bm{k}}\delta b^z_{\gamma\bm{k}}(t) \notag
\nonumber \\
&\quad
+2\Delta^{\mathrm{eq}}_\gamma\delta\sigma^z_{\gamma\bm{k}}(t)+2\sigma^{z,\mathrm{eq}}_{\gamma\bm{k}}\delta\Delta'_\gamma(t), \\
\label{z_eqn}
\partial_t\delta\sigma^z_{\gamma\bm{k}}(t)&=-2\Delta^{\mathrm{eq}}_\gamma\delta\sigma^y_{\gamma\bm{k}}(t)+2\sigma^{x,\mathrm{eq}}_{\gamma\bm{k}}\delta\Delta''_\gamma(t), 
\end{align}
where we have defined the deviations, 
$\delta\bm{\sigma}_{\gamma\bm{k}}(t)=\bm{\sigma}_{\gamma\bm{k}}(t)-\bm{\sigma}^{\mathrm{eq}}_{\gamma\bm{k}}$, $\delta\Delta'_\gamma(t)=\Delta'_\gamma(t)-\Delta^{\mathrm{eq}}_\gamma$, $\delta\Delta''_\gamma(t)=\Delta''_\gamma(t)-0$, and the effect of the laser, $\delta b^z_{\gamma\bm{k}}(t)=b^z_{\gamma\bm{k}}(t)-\epsilon_{\gamma\bm{k}}\simeq(e^2/2)\sum_{ij}(\partial_{k_i}\partial_{k_j}\epsilon_{\gamma\bm{k}})A_i(t)A_j(t)$.
Fourier transforms, e.g. $\delta\Delta'_\gamma(t)=\int_{-\infty}^\infty \frac{d\omega}{2\pi}\delta\Delta'_\gamma(\omega)e^{i\omega t}$, give
\begin{align}
\left(
\begin{array}{c}
\delta\sigma^x_{\gamma\bm{k}}(\omega) \\ \delta\sigma^y_{\gamma\bm{k}}(\omega) \\ \delta\sigma^z_{\gamma\bm{k}}(\omega)
\end{array}
\right)&=-\left(
\begin{array}{ccc}
4\epsilon_{\gamma\bm{k}}^2 & 2i\omega\epsilon_{\gamma\bm{k}} & 4\Delta_\gamma\epsilon_{\gamma\bm{k}} \\
-2i\omega\epsilon_{\gamma\bm{k}} & 4E^2_{\gamma\bm{k}} & -2i\omega\Delta_\gamma \\
4\Delta_\gamma\epsilon_{\gamma\bm{k}} & 2i\omega\Delta_\gamma & 4\Delta^2_\gamma
\end{array}
\right) \notag \\
&\quad\times\frac{\sigma^x_{\gamma\bm{k}}}{\Delta_\gamma(4E^2_{\gamma\bm{k}}-\omega^2)}\left(
\begin{array}{c}
-\delta\Delta'_\gamma(\omega) \\ -\delta\Delta''_\gamma(\omega) \\ \delta b^z_{\gamma\bm{k}}(\omega)
\end{array}
\right),
\label{linearized}
\end{align}
with the superscript ``eq" dropped from $\Delta_\gamma, \sigma^x_{\gamma\bm{k}}$. 

We consider a linearly-polarized light, and define $x$ axis parallel to the polarization direction. Then the electric field can be described by $\bm{A}(t)=A(t)\hat{\bm{x}}$ ($\hat{\bm{x}}$: a unit vector along $x$ axis)
giving
\begin{equation}
\delta b^z_{\gamma\bm{k}}(\omega)=\frac{e^2}{2}A^2(\omega)\frac{\partial^2\epsilon_{\gamma\bm{k}}}{\partial k_x^2},
\label{driving_force}
\end{equation}
where $A^2(\omega)$ is the Fourier transform of $A(t)^2$. 
The above pseudospin formulation can thus be regarded as a linear-response theory to a ``nonlinear'' field $A(t)^2$. In the following we shall solve the linearized Eq.(\ref{linearized}) self-consistently.

\section{Optical excitation of\\ 
Leggett modes} \label{Leggett_mode}

First we derive the solution for the imaginary parts of the gaps. Summing $\delta\sigma^y_{\gamma\bm{k}}(\omega)$ with Eq.(\ref{linearized}) and using the self-consistent constraint Eq.(\ref{self_consistency}), we obtain
\begin{widetext}
\begin{align}
\left(
\begin{array}{c}
\delta\Delta''_\alpha(\omega) \\ \delta\Delta''_\beta(\omega)
\end{array}
\right) = &\frac{e^2A^2(\omega)}{i\omega\left[\omega^2F_\alpha(\omega)F_\beta(\omega)\det\lambda
+\lambda_{\beta\alpha}\Delta_\alpha F_\alpha(\omega)+\lambda_{\alpha\beta}\Delta_\beta F_\beta(\omega)\right]} \notag \\
&\times\left(
\begin{array}{cc}
\omega^2F_\beta(\omega)\det\lambda+\lambda_{\beta\alpha}\Delta_\alpha & \lambda_{\alpha\beta}\Delta_\alpha \\
\lambda_{\beta\alpha}\Delta_\beta & \omega^2F_\alpha(\omega)\det\lambda+\lambda_{\alpha\beta}\Delta_\beta
\end{array}
\right)\left(
\begin{array}{c}
\Delta_\alpha Y_\alpha(\omega) \\ \Delta_\beta Y_\beta(\omega)
\end{array}\right),
\label{solution}
\end{align}
\end{widetext}
where
\begin{align}
F_\gamma(\omega)&=\frac{1}{D_\gamma}\sum_{\bm{k}}\frac{\sigma^x_{\gamma\bm{k}}}{4E_{\gamma\bm{k}}^2-\omega^2}, 
\label{function_F2} \\
Y_\gamma(\omega)&=\frac{1}{D_\gamma}\sum_{\bm{k}}\frac{\sigma^x_{\gamma\bm{k}}}{4E_{\gamma\bm{k}}^2-\omega^2}\frac{\partial^2\epsilon_{\gamma\bm{k}}}{\partial k_x^2}, \label{function_Y2} \\
\lambda_{\gamma\gamma'}=V_{\gamma\gamma'}&D_{\gamma'},~
\det\lambda=\lambda_{\alpha\alpha}\lambda_{\beta\beta}-\lambda_{\alpha\beta}\lambda_{\beta\alpha}, \label{dlpi}
\end{align}
with 
\begin{equation}
D_\gamma=\sum_{\bm{k}}\delta(\epsilon-\epsilon_{\gamma\bm{k}}) \label{DOS}
\end{equation}
being the density of states on the Fermi surface of $\gamma$-band, 
assumed to be constant around the Fermi energy ($\epsilon\approx0$).
We summarize the definition of symbols such as Eqs.(\ref{function_F2}-\ref{DOS}) in Appendix A.

If we replace the $\bm{k}$ summation with an energy integral, 
\begin{equation}
\sum_{\bm{k}}=\int d\epsilon\sum_{\bm{k}}\delta(\epsilon-\epsilon_{\gamma\bm{k}}), \label{replacement}
\end{equation}
we obtain
\begin{eqnarray}
F_\gamma(\omega) &=& \int_{-\infty}^{\infty}d\epsilon
\frac{\Delta_\gamma\tanh\left(\sqrt{\epsilon^2+\Delta_\gamma^2}/2k_BT\right)}
{2\sqrt{\epsilon^2+\Delta_\gamma^2}\left(4\epsilon^2+4\Delta_\gamma^2-\omega^2\right)}, \label{function_F} \\
Y_\gamma(\omega) &=& c_{\gamma0}F_\gamma(\omega), \label{function_Y}
\end{eqnarray}
where the coefficient $c_{\gamma0}$ is defined by a series expansion,
\begin{equation}
\sum_{\bm{k}}\delta(\epsilon-\epsilon_{\gamma\bm{k}})\frac{\partial^2\epsilon_{\gamma\bm{k}}}{\partial k_x^2}
=D_\gamma(c_{\gamma0}+c_{\gamma1}\epsilon+c_{\gamma2}\epsilon^2\cdots). \label{expansion}
\end{equation}
Since we are interested in low-energy responses of superconductors, 
the relevant excitations are those 
with the energy scale $\sim\Delta$,
far below the bandwidth in the weak-coupling regime.
Thus the higher-order terms in the above expansion are of less importance, 
so that we neglect the $c_{\gamma n}$ terms with $n\ge2$. 
Physical meaning of $c_{\gamma0}$ and $c_{\gamma1}$ retained here are 
simple: 
consider a parabolic band with an isotropic effective mass, 
then $c_{\gamma0}$ is the inverse effective mass, 
while $c_{\gamma1}$ vanishes.
Therefore, one can roughly say that $c_{\gamma0}$ and $c_{\gamma1}$ measure 
the effective mass 
and nonparabolicity of an energy band, respectively.

The formalism presented here is general enough to be applicable to any band structures.
It also describes the polarization dependence of the optical response, 
since the coefficients $c_{\gamma n}$ depend on the relative 
angles between $x$ axis (polarization direction of the incident light) and the crystallographic axes.

In the linearized Eqs.(\ref{x_eqn}-\ref{z_eqn}), the imaginary part of $\Delta_\gamma$ is proportional to
the phase $\theta_\gamma$ defined by $\Delta_\gamma = |\Delta_\gamma|e^{i\theta_\gamma}$. 
The phase difference between the two gaps is a physical (gauge-invariant) quantity, while each phase is not. 
Motion of the phase difference is governed by
\begin{equation}
\delta[\theta_\alpha(\omega)-\theta_\beta(\omega)]=\frac{\delta\Delta''_\alpha(\omega)}{\Delta_\alpha}-\frac{\delta\Delta''_\beta(\omega)}{\Delta_\beta}=-e^2A^2(\omega)i\omega L(\omega), \label{Leggett_solution}
\end{equation}
where
\begin{align}
&L(\omega) \notag \\ 
&=\frac{(c_{\alpha0}-c_{\beta0})F_\alpha(\omega)F_\beta(\omega)\det\lambda}{\omega^2F_\alpha(\omega)F_\beta(\omega)\det\lambda+\lambda_{\beta\alpha}\Delta_\alpha F_\alpha(\omega)
+\lambda_{\alpha\beta}\Delta_\beta F_\beta(\omega)}. 
\label{L_factor}
\end{align}
This solution describes a 
\textit{resonance between the squared electric field and the Leggett mode}, whose energy is determined by 
the frequency at which the denominator of Eq.(\ref{L_factor}) vanishes\cite{Blumberg},
\begin{equation}
\omega^2F_\alpha(\omega)F_\beta(\omega)\det\lambda+\lambda_{\beta\alpha}\Delta_\alpha F_\alpha(\omega)+\lambda_{\alpha\beta}\Delta_\beta F_\beta(\omega)=0. \label{Leggett_energy}
\end{equation}

For weak enough $V_{\alpha\beta}$, the solution of this equation can be approximately given by
\begin{equation}
\omega^2=\omega_{\mathrm{L}}^2\equiv-4\left(\frac{\lambda_{\alpha\beta}+\lambda_{\beta\alpha}}
{\det\lambda}\right)\Delta_\alpha\Delta_\beta 
\label{Leggett_frequency}
\end{equation}
at $T=0$, where $F_\gamma(\omega)$ reduces to
\begin{align}
F_\gamma(\omega)&=\frac{\Delta_\gamma}{\omega\sqrt{4\Delta_\gamma^2-\omega^2}}
\sin^{-1}\left(\frac{\omega}{2|\Delta_\gamma|}\right).
\label{simplest_F}
\end{align}
The right-hand side of Eq.(\ref{Leggett_frequency}) is positive-definite
for $\det\lambda>0$, 
because $V_{\alpha\beta}$ and $\Delta_\alpha\Delta_\beta$ necessarily have opposite signs (see the previous section).
The mode energy (\ref{Leggett_frequency}) was originally derived by Leggett\cite{Leggett}.
We consider a monochromatic wave turned on at $t=0$, $A(t>0)=A_0\sin\Omega t$, 
for which the Fourier transform of the squared vector potential is given by
\begin{equation}
A^2(\omega)=\frac{4A_0^2\Omega^2}{i\omega\left(4\Omega^2-\omega^2\right)},
\end{equation}
where $\omega$ on the right-hand side stands for $\omega-i0$.
Then, for small $V_{\alpha\beta}$,
we can show that approximately
\begin{equation}
\delta[\theta_\alpha-\theta_\beta] \simeq 
\frac{4(c_{\alpha0}-c_{\beta0})\Omega^2e^2A_0^2}{(\omega^2-4\Omega^2)(\omega^2-\omega_{\mathrm{L}}^2)}.
\label{resonanceshape}
\end{equation} 
An inverse Fourier transform gives the temporal behavior,
\begin{equation}
\delta[\theta_\alpha-\theta_\beta] \simeq 
\frac{4(c_{\alpha0}-c_{\beta0})\Omega^2e^2A_0^2}{4\Omega^2-\omega_{\mathrm{L}}^2}
\left(\frac{\sin\omega_{\mathrm{L}}t}{\omega_{\mathrm{L}}}-\frac{\sin2\Omega t}{2\Omega}\right),
\label{resonantmotion}
\end{equation} 
for $t>0$.  
When $2\Omega$ (the incident wave frequency doubled) 
is close to $\omega_{\mathrm{L}}$, 
the poles ($\omega=\pm 2\Omega, \pm \omega_{\mathrm{L}}$)
on the right-hand side of Eq.(\ref{resonanceshape}) merge, 
leading to a resonance between the Leggett mode and forced oscillation due to the electromagnetic wave. In the time domain, this appears as a factor $(4\Omega^2-\omega_{\mathrm{L}}^2)$ in the denominator of Eq.(\ref{resonantmotion}), which enhances both the forced oscillation and the excited Leggett mode. Under the exact resonance condition, $2\Omega=\omega_{\mathrm{L}}$, Eq.(\ref{resonanceshape}) gives
\begin{equation}
\delta[\theta_\alpha-\theta_\beta] \simeq 
\frac{(c_{\alpha0}-c_{\beta0})\omega^2_{\mathrm{L}}e^2A_0^2}{2}
\left(\frac{\sin\omega_{\mathrm{L}}t}{\omega_{\mathrm{L}}}-t\cos\omega_{\mathrm{L}}t\right),
\end{equation}
whose amplitude diverges for $t\to\infty$ (so that, strictly speaking, the linearized equation of motion fails right at the resonance and nonlinear effects will dominate in the long-time behavior).  
This provides a new concept of \textit{resonant excitation of Leggett mode}.

\begin{figure}
	\begin{center}
		\includegraphics[width=8cm,clip]{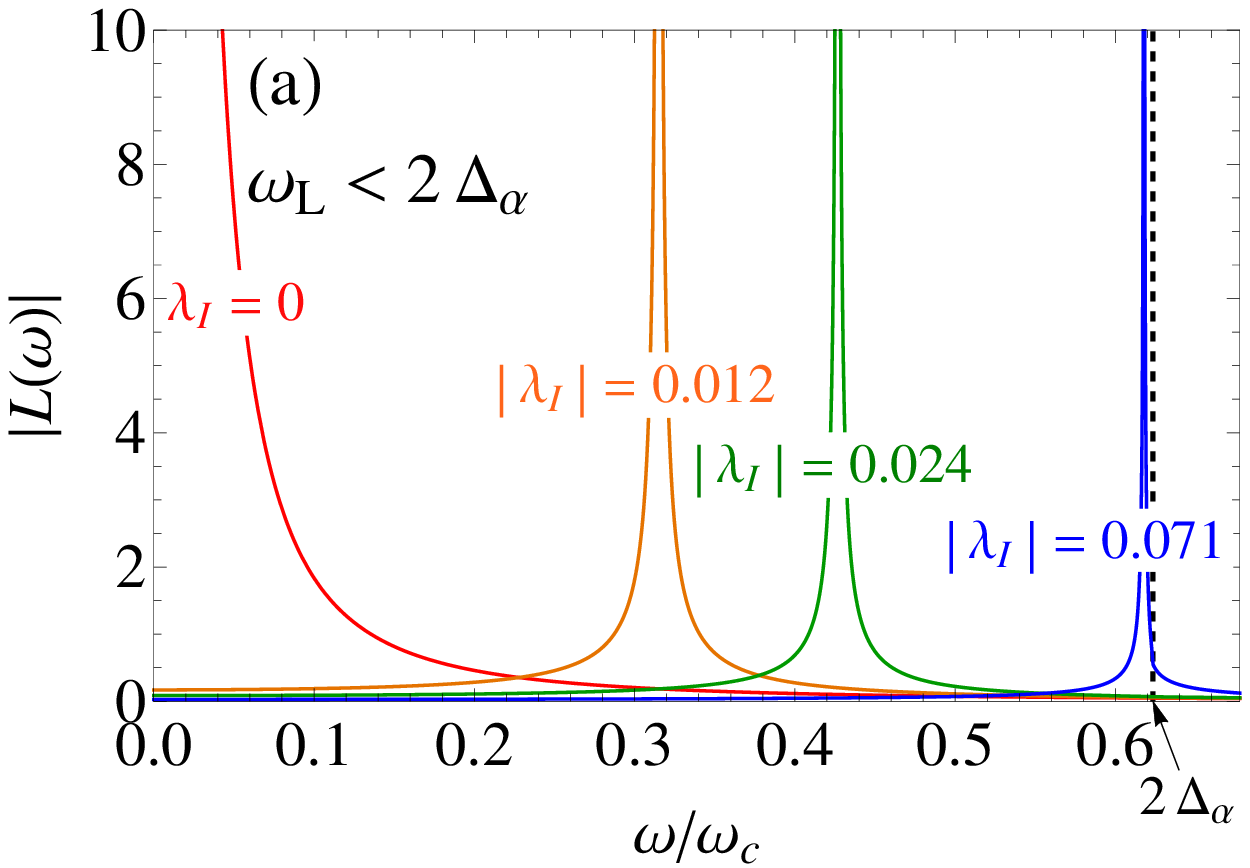}
		\includegraphics[width=8cm,clip]{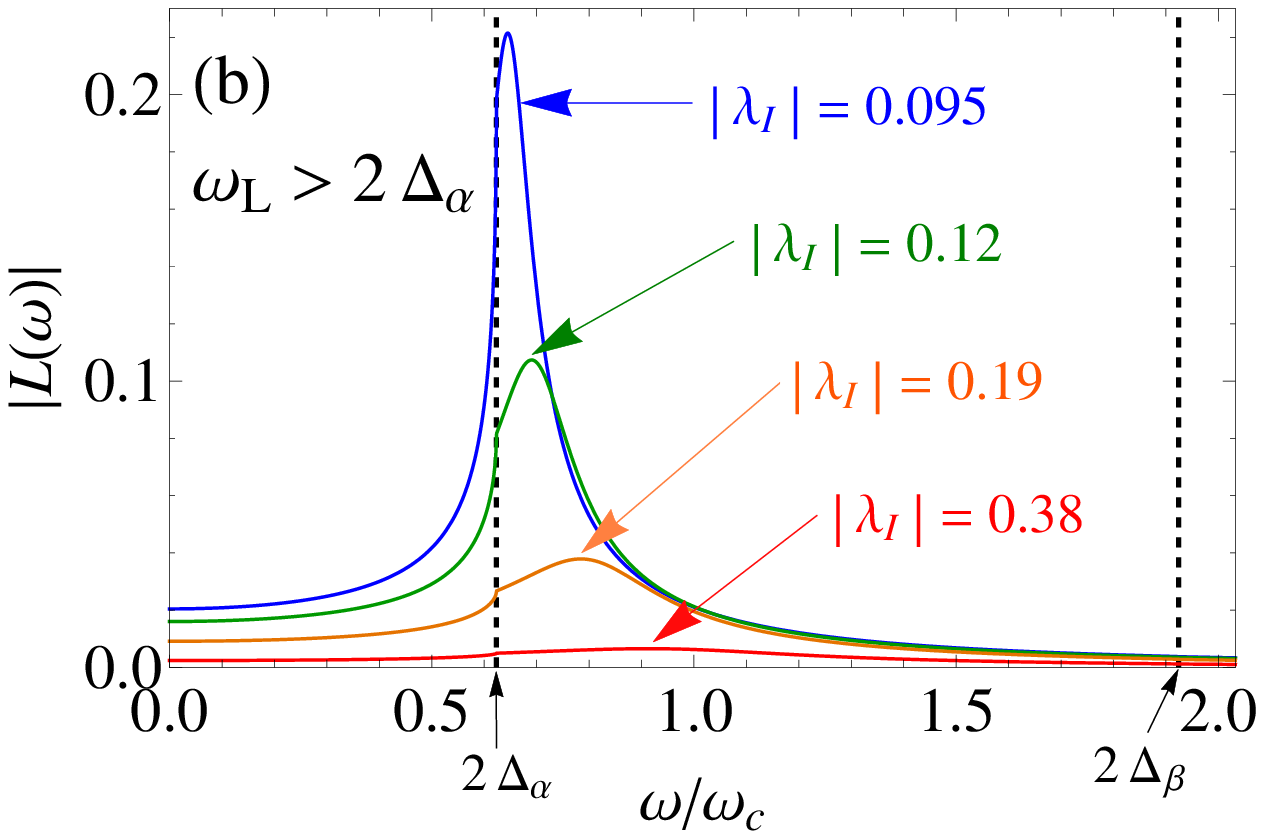}
				
		\caption{$|L(\omega)|$ at $T=0$ for several 
values of the dimensionless interband pairing interaction 
$|\lambda_I|$ varied from 0 to $0.071$ (a)
and from $0.095$ to $0.38$ (b), for 
$\lambda_{\alpha\alpha}=-0.28,~\lambda_{\beta\beta}=-0.96,~\lambda_{\alpha\beta}/\lambda_{\beta\alpha}=0.73,~\Delta_\alpha/\omega_c=0.31,~\Delta_\beta/\omega_c=0.96,$ and $c_{\alpha0}-c_{\beta0}=1$. Position of the peak defines $\omega_{\mathrm{L}}$.} \label{L_factor_fig}
	\end{center}
\end{figure}

When $V_{\alpha\beta}$ is not small, the above approximate solution cannot be used. 
To study the effects of increasing $V_{\alpha\beta}$ on the Leggett mode resonance, 
we examine the spectral feature of $L(\omega)$, 
which can be regarded as the resonance factor for the Leggett mode.
As a measure of the interband coupling strength, we define a dimensionless quantity,
\begin{equation}
\lambda_I \equiv V_{\alpha\beta}\sqrt{D_\alpha D_\beta}, \label{dlipi}
\end{equation}
with $\lambda_I^2=\lambda_{\alpha\beta}\lambda_{\beta\alpha}$.
Note that $L(\omega)$ remains the same for positive and negative $\lambda_I$,
because, when the sign of $\lambda_I$ (i.e., that of $V_{\alpha\beta}$) is inverted, the sign of $\Delta_\alpha\Delta_\beta$ is also changed (see the previous section), 
and these sign changes are canceled between the denominator and numerator in Eq.(\ref{L_factor}).
Figure \ref{L_factor_fig} illustrates the absolute value of $L(\omega)$ 
for various values of $\lambda_I$ at $T=0$.
The parameters are taken to be $c_{\alpha0}-c_{\beta0}=1$, $\lambda_{\alpha\alpha}=-0.28,~\lambda_{\beta\beta}=-0.96,~\lambda_{\alpha\beta}/\lambda_{\beta\alpha}=D_\beta/D_\alpha=0.73,~\Delta_\alpha/\omega_c=0.31,~\Delta_\beta/\omega_c=0.96$, which are 
chosen for $\mathrm{MgB_2}$\cite{Liu, Iavarone, Blumberg}, 
where the interband interaction $\lambda_I=-0.19$ 
is estimated from Ref.\onlinecite{Liu}.
Energy is measured in units of the cut-off energy $\omega_c$, which is necessary to obtain a finite solution of the BCS gap equation.
Here we intended to look into the peak positions relative to the 
gap function, so that we show the result when the modification of 
$\Delta_\gamma$ by changing $\lambda_I$ through the gap equation is ignored.  We have qualitatively similar behavior 
of $\left|L(\omega)\right|$ (peak widths, etc.) when we take account of that.

\begin{figure}
	\begin{center}
		\includegraphics[width=8cm,clip]{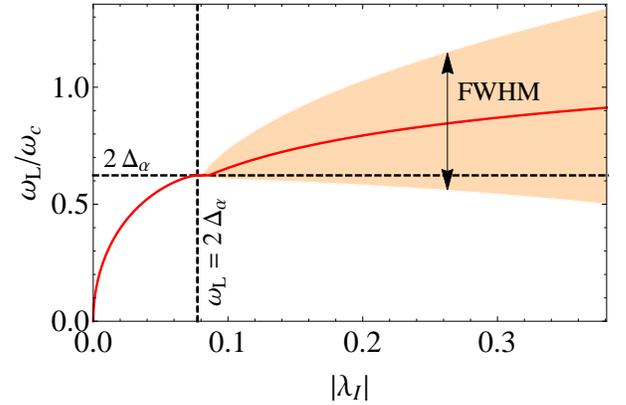}
		\caption{The energy of the Leggett mode 
 (red curve) and halfwidth (full width at half maximum; FWHM) of $|L(\omega)|$ (orange shading) 
against the interband pairing interaction $|\lambda_I|$.  The horizontal dashed line indicates $2\Delta_\alpha$ (the smaller of the two gaps), while the vertical one the critical value of $|\lambda_I|$ at which 
$\omega_{\mathrm{L}}$ reaches $2\Delta_\alpha$. Values of the used parameters are the same as in Fig. \ref{L_factor_fig}.} \label{fig_energy}
	\end{center}
\end{figure}

When the interband coupling is relatively small,
Eq.(\ref{Leggett_energy}) has a real solution smaller than $2\Delta_\alpha$ so that
$|L(\omega)|$ diverges, as shown in Fig. \ref{L_factor_fig}(a). 
This contrasts with the case of the relatively strong interband coupling, 
where Eq.(\ref{Leggett_energy}) has no real solution and thus $|L(\omega)|$ does not diverge;
instead, $|L(\omega)|$ is peaked at a frequency above $2\Delta_\alpha$, as shown in Fig. \ref{L_factor_fig}(b).
We define the peak frequency as the energy of the Leggett mode, $\omega_{\mathrm{L}}$, and
plot $\omega_{\mathrm{L}}$ and the halfwidth of $|L(\omega)|$ against 
$|\lambda_I|$ in Fig. \ref{fig_energy}.
For small $|\lambda_I|$, $|L(\omega)|$ diverges 
at $\omega=\omega_{\mathrm{L}}$ as determined by the solution of Eq.(\ref{Leggett_energy}), so that the halfwidth is ill-defined (or zero).  
By contrast, when $\omega_{\mathrm{L}}$ exceeds the smaller gap $2\Delta_\alpha$ 
as $\lambda_I$ is increased, 
$|L(\omega)|$ stops diverging and starts to have finite widths, 
where both $\omega_{\mathrm{L}}$ and the halfwidth increase monotonically with $|\lambda_I|$. Since the lifetime of the mode is roughly given by the inverse of the halfwidth, one can say that the Leggett mode is a long-lived mode only when its energy is below the superconducting gaps. 
This is due to suppression of the decay from the Leggett mode to lower-energy quasi-particles\cite{Klein}.

Even when the lifetime of the Leggett mode is finite, the peak of $|L(\omega)|$ at $\omega=\omega_{\mathrm{L}}$ and the pole of $A^2(\omega)$ at $\omega=2\Omega$ can constructively enhance the optical response in Eq.(\ref{Leggett_solution}). Thus a resonant excitation of the Leggett mode is also available for short-lived cases, while sharpness of the resonance will be degraded.

The mechanism of the \textit{light-induced Leggett mode} found here 
can be explained as follows. 
In the absence of $V_{\alpha\beta}$, Eq.(\ref{solution}) drives a rotation of the 
phases in the form of
\begin{equation}
\Delta_\gamma(t)=\Delta_\gamma(0)\exp\left(ic_{\gamma0}e^2\int_0^tdt'A(t')^2\right). \label{phase}
\end{equation}
This implies that a Cooper pair with charge $2e$ in $\gamma$-band feels an effective electrostatic potential $\Phi_{\gamma}^{\mathrm{eff}}=(c_{\gamma0}/2)eA(t)^2$. In single-band superconductors, this phase can be gauged out\cite{Tsuji}, and the effective potential has no physical meaning. By contrast, 
a two-band system has two phase variables, which cannot be simultaneously gauged out: 
the phase difference is gauge-invariant. Correspondingly, a difference in the effective potential between the two bands has a physical effect: when $c_{\alpha0}\neq c_{\beta0}$, an effective voltage emerges between the two bands, leading to a phase difference between the gaps. 
Such a phase difference is not favored in the presence of $\lambda_I\neq0$, 
because a particular relation ($s_{\pm}$ or $s_{++}$) is imposed upon the ground state.
Therefore, the interband Josephson coupling $\lambda_I$ produces a restoring force for the phase difference, which 
acts to induce the Leggett mode.\\

\section{Optical excitation of Higgs modes} \label{Higgs_modes}

Now we move on to the real parts of the gaps.
The self-consistent solution of Eq.(\ref{linearized}) is given by
\begin{equation}
\left(
\begin{array}{c}
\delta\Delta'_\alpha(\omega) \\ \delta\Delta'_\beta(\omega)
\end{array}\right)=\frac{e^2A^2(\omega)}{2}\left(
\begin{array}{c}
H_\alpha(\omega) \\ H_\beta(\omega)
\end{array}\right),
\label{H_solution}
\end{equation}
where we have defined
\begin{widetext}
\begin{align}
\left(
\begin{array}{c}
H_\alpha(\omega) \\ H_\beta(\omega)
\end{array}\right) = &-\frac{1}{G_\alpha(\omega)G_\beta(\omega)\det\lambda-\lambda_{\beta\alpha}\Delta_\alpha G_\alpha(\omega)-\lambda_{\alpha\beta}\Delta_\beta G_\beta(\omega)}
\notag \\
&\times\left(
\begin{array}{cc}
G_\beta(\omega)\det\lambda-\lambda_{\beta\alpha}\Delta_\alpha & -\lambda_{\alpha\beta}\Delta_\alpha \\
-\lambda_{\beta\alpha}\Delta_\beta & G_\alpha(\omega)\det\lambda-\lambda_{\alpha\beta}\Delta_\beta
\end{array}\right)
\left(
\begin{array}{c}
\Delta_\alpha X_\alpha(\omega) \\
\Delta_\beta X_\beta(\omega)
\end{array}\right), 
\label{H_factor}
\end{align}
\end{widetext}
\begin{align}
G_\gamma(\omega)&=(4\Delta_\gamma^2-\omega^2)F_\gamma(\omega), \label{function_G} \\
X_\gamma(\omega)&=\frac{1}{D_\gamma}\sum_{\bm{k}}\frac{4\sigma^x_{\gamma\bm{k}}\epsilon_{\gamma\bm{k}}}{4E_{\gamma\bm{k}}^2-\omega^2}\frac{\partial^2\epsilon_{\gamma\bm{k}}}{\partial k_x^2} \notag \\
&=-c_{\gamma1}\left[\frac{\lambda_{\overline{\gamma\gamma}}\Delta_\gamma-\lambda_{\gamma\overline{\gamma}}\Delta_{\overline{\gamma}}}{\det\lambda}+G_\gamma(\omega)\right],
\label{function_X}
\end{align}
with $c_{\gamma1}$ defined by Eq.(\ref{expansion}), and $\overline{\alpha}=\beta$, $\overline{\beta}=\alpha$.
The function $H_\gamma(\omega)$ $(\gamma=\alpha, \beta)$ describes resonance between the squared electric field and the Higgs modes, where $X_\gamma(\omega)$ gives rise to coupling between them.

\begin{figure}
	\begin{center}
		\includegraphics[width=8cm,clip]{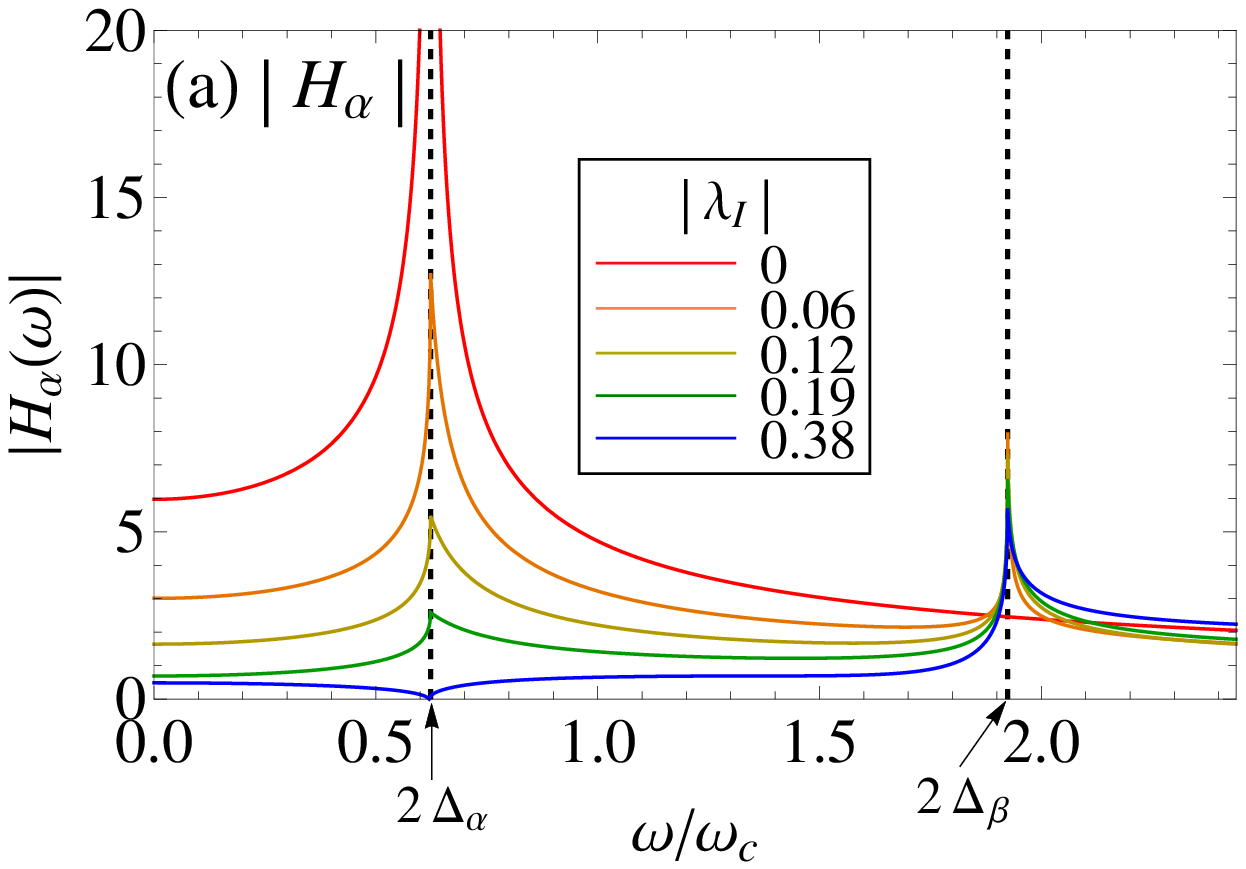}
		\includegraphics[width=8cm,clip]{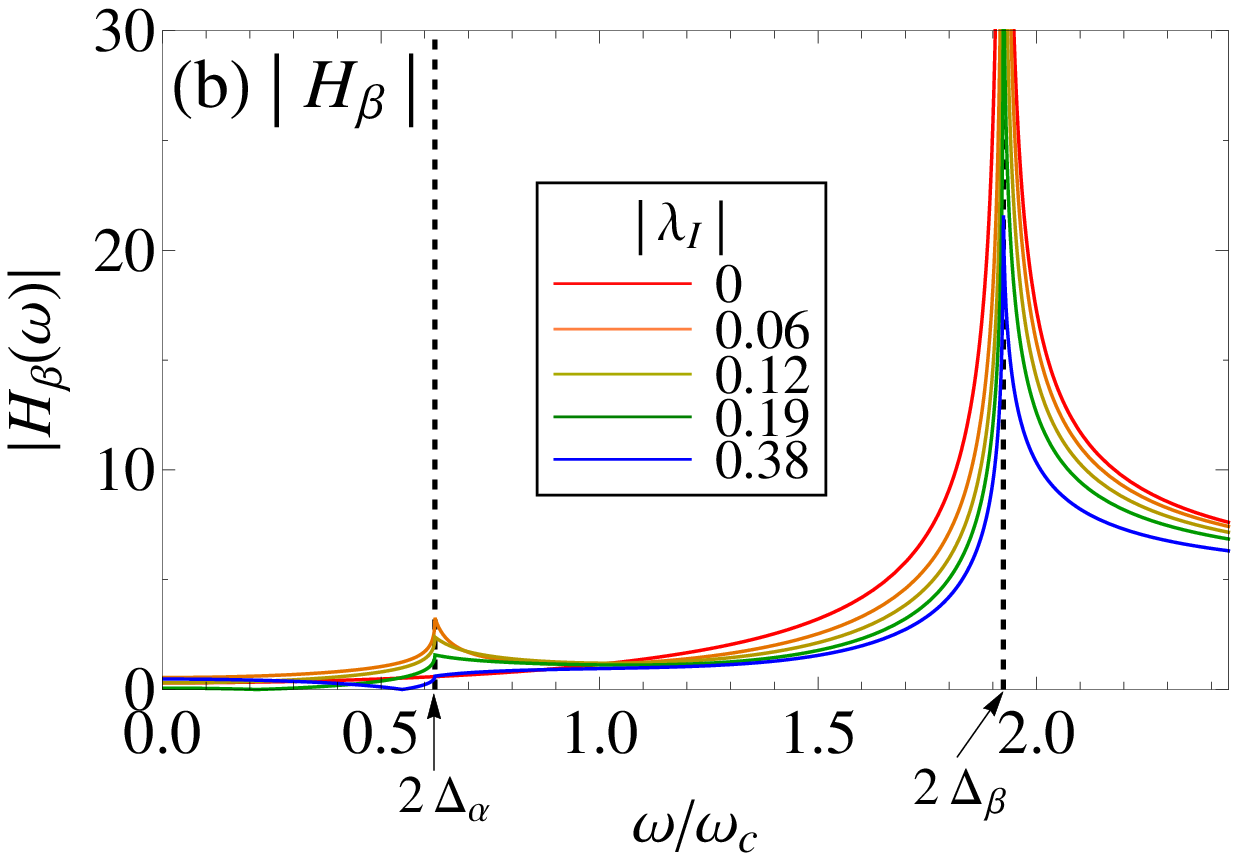}

		\caption{$|H_\alpha(\omega)|$ (a) and $|H_\beta(\omega)|$ (b) for several values of $|\lambda_I|$ varied from 0 to 0.38, with $c_{\alpha1}=c_{\beta1}=-1$ and the same parameters as Fig. \ref{L_factor_fig}.} \label{H_factor_fig}
	\end{center}
\end{figure}

\begin{figure}
	\begin{center}
		\includegraphics[width=8.5cm,clip]{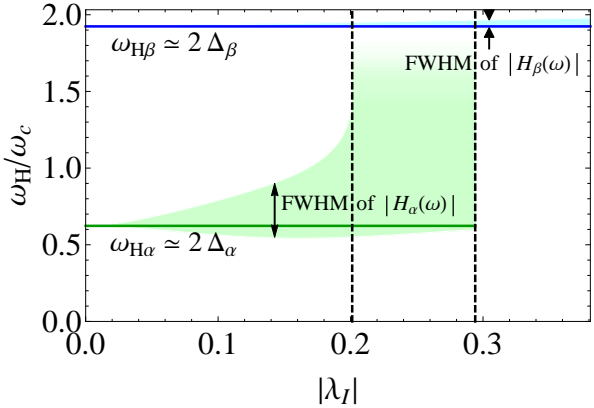}
		\caption{Dependence of the Higgs modes on the 
interband pairing interaction $|\lambda_I|$. Green (blue) line represents the mode energy $\omega_{\mathrm{H}\alpha}$ ($\omega_{\mathrm{H}\beta}$), defined as the peak of $|H_\alpha(\omega)|$ ($|H_\beta(\omega)|$), associated with 
the smaller gap $2\Delta_\alpha$ (larger gap $2\Delta_\beta$). The halfwidths are indicated by green (blue) regions. Left dashed line indicates 
$|\lambda_I|$ at which the upper halfwidth of $|H_\alpha(\omega)|$ diverges, while the right one the 
$|\lambda_I|$ at which the peak of $|H_\alpha(\omega)|$ at $\omega=2\Delta_\alpha$ disappears.
Used parameters are the same as Fig. \ref{H_factor_fig}.} \label{fig_energy_H}
	\end{center}
\end{figure}

We plot $\left|H_\gamma(\omega)\right|$ in Fig. \ref{H_factor_fig}, for $c_{\alpha1}=c_{\beta1}=-1$ and the parameters estimated for $\mathrm{MgB_2}$ as in Fig. \ref{L_factor_fig}.
Here, too, we have shown the result when the modification of 
$\Delta_\gamma$ by changing $\lambda_I$ is ignored, while qualitatively similar behavior is obtained 
when we take account of that.
The peaks at $\omega=2\Delta_\alpha,~2\Delta_\beta$ (which we call $\omega_{\mathrm{H}\alpha}$ and $\omega_{\mathrm{H}\beta}$, respectively) represent the Higgs modes.
Due to the interband interaction, they are coupled to each other, so that both of $|H_\alpha(\omega)|$ and $|H_\beta(\omega)|$ exhibit two peaks 
at $2\Delta_{\alpha}$ and $2\Delta_{\beta}$ 
for nonzero $\lambda_I$ (while they reduce to single Higgs modes in the limit $\lambda_I\to0$).
As in the single-band case\cite{Tsuji,Matsunaga2} and in the Leggett mode discussed in the previous section, 
the Higgs modes in two-band superconductors can be resonantly excited by electromagnetic waves, now at $2\Omega \simeq \omega_{\mathrm{H}\alpha},~\omega_{\mathrm{H}\beta}$.
The two Higgs modes, however, show very different resonance features. 
$|H_\beta(\omega)|$ for the larger gap has a sharp resonance peak at $\omega=2\Delta_\beta$ for all values of $\lambda_I$ studied here,
although the peak height somewhat decreases with increasing  $\lambda_I$.
By contrast, the peak of $|H_\alpha(\omega)|$ at $\omega=2\Delta_\alpha$ 
for the smaller gap is
rapidly suppressed and broadened with increasing $|\lambda_I|$, and finally disappears [Fig. \ref{H_factor_fig}(a), blue curve].
We summarize the peak positions and widths of $|H_\gamma(\omega)|$ in Fig. \ref{fig_energy_H}. 
Unlike the relatively narrow peak of $|H_\beta(\omega)|$ around $\omega=2\Delta_\beta$,
the peak of $|H_\alpha(\omega)|$ at $\omega=2\Delta_\alpha$ 
becomes rapidly broadened as $|\lambda_I|$ is increased, and 
when $|\lambda_I|$ exceeds a certain value, the ``width'' can no longer be defined.
If we further increase $|\lambda_I|$, the peak vanishes at a certain point.
This indicates that the Higgs mode associated with the larger gap is more stable than the one associated with the lower gap, 
contrary to a naive 
expectation that a lower-energy excitation would be more stable.

The reason why the Higgs mode with the lower energy disappears can be explained as follows.  
For small $|\lambda_I|$, the two condensates, hence 
the two Higgs modes, are almost independent of each other.
For strong enough $|\lambda_I|$, 
on the other hand, the two condensates are so strongly coupled that they can be regarded as almost one condensate. Its character is dominated by the component 
having the larger superfluid density, hence the larger gap. Therefore, only one Higgs mode with the higher energy survives for large enough interband interactions.

\section{Temperature effects on\\ 
mode resonances}
\begin{figure}
	\begin{center}
		\includegraphics[width=7.6cm,clip]{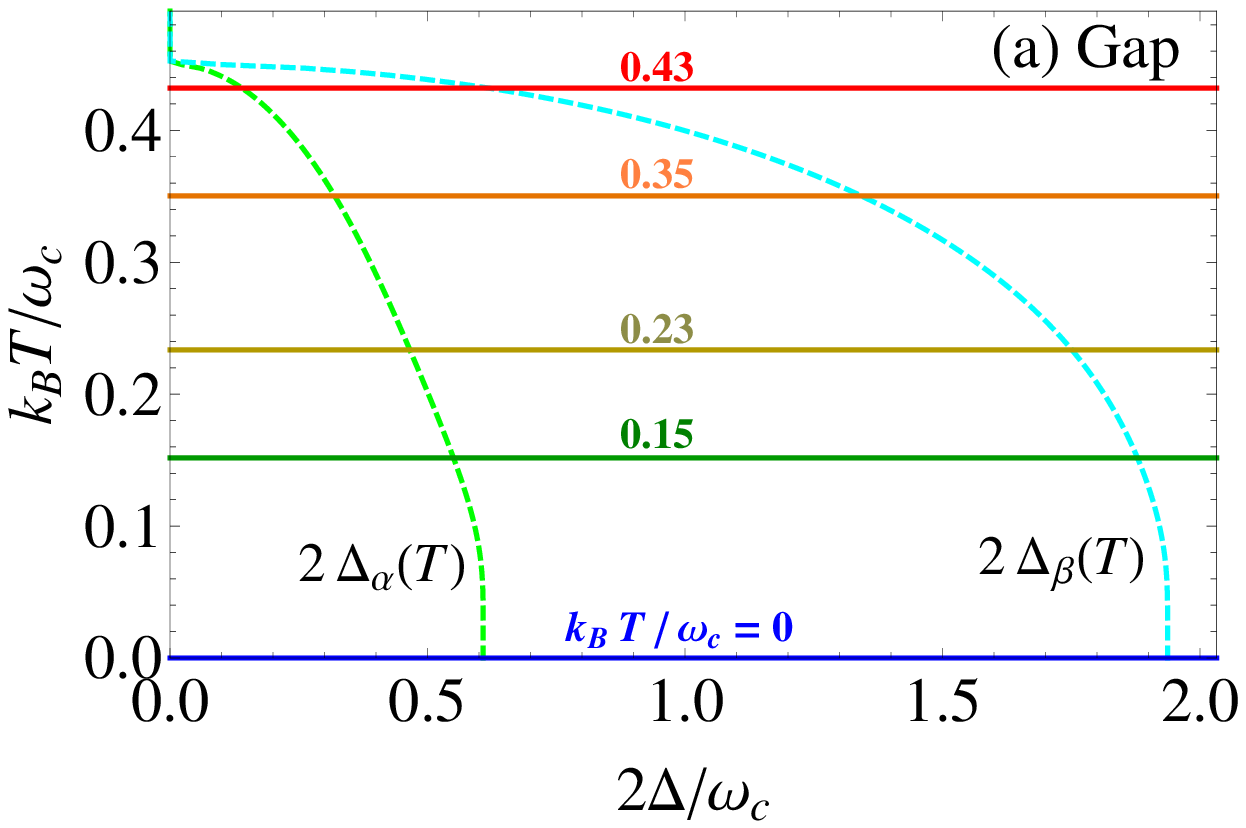}
		\includegraphics[width=7.8cm,clip]{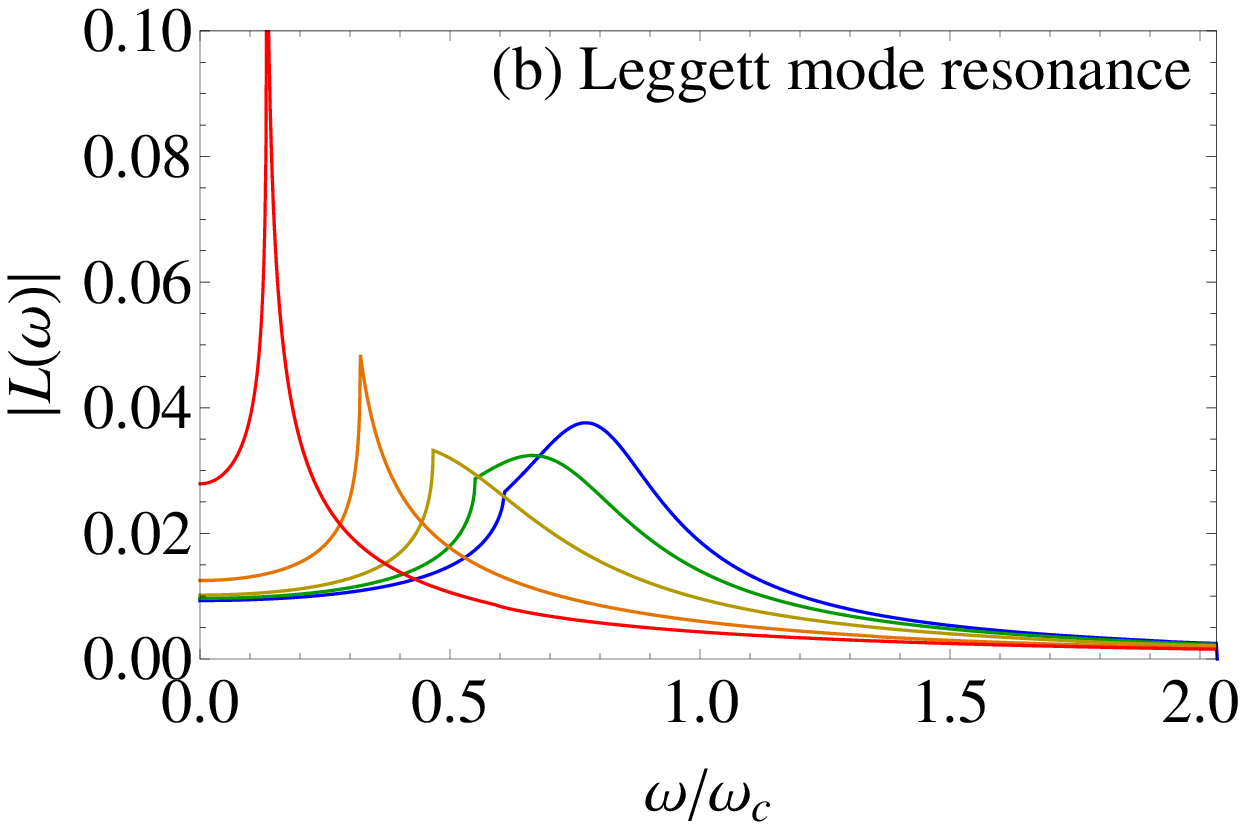}
		\includegraphics[width=7.6cm,clip]{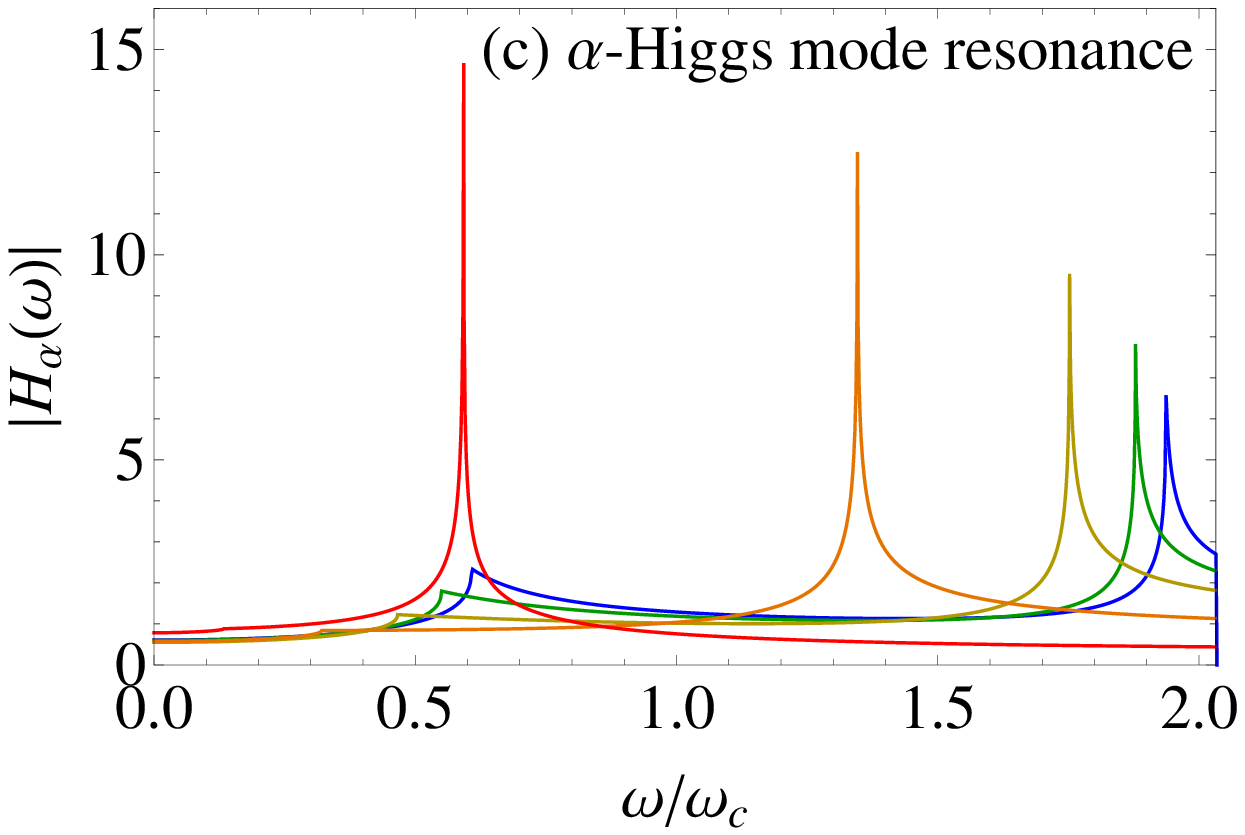}
		\includegraphics[width=7.6cm,clip]{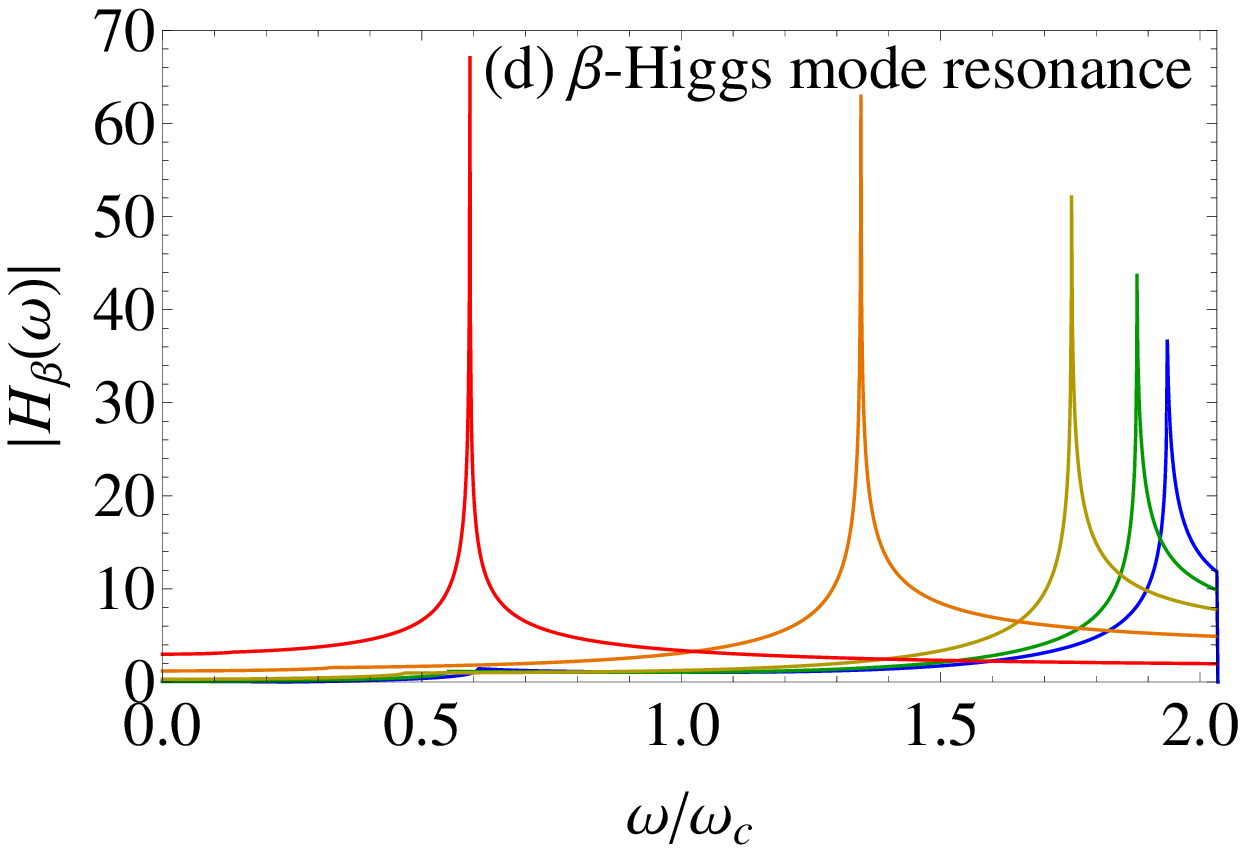}

		\caption{Dependence of the gaps (horizontal axis) 
$2\Delta_\alpha(T)$ and $2\Delta_\beta(T)$ (green and blue dashed lines, respectively) against temperature (vertical axis), for $\lambda_{\alpha\alpha}=-0.28,~\lambda_{\beta\beta}=-0.96,~\lambda_{\alpha\beta}/\lambda_{\beta\alpha}=0.73$ and $\lambda_I=-0.19$ (a).  Temperatures 
chosen in panels (b-d) are indicated by horizontal lines: $k_BT/\omega_c=0$ (blue), 0.15 (green), 0.23 (khaki), 0.35 (orange), and 0.43 (red).
$|L(\omega)|$ (b), $|H_\alpha(\omega)|$ (c) and $|H_\beta(\omega)|$ (d) for several values of $T$ with $c_{\alpha0}-c_{\beta0}=1$ and $c_{\alpha1}=c_{\beta1}=-1$.}\label{temp_dep}
	\end{center}
\end{figure}

The effect of temperature on the sharpness of resonances is an important issue, especially from experimental points of view. 
It is also directly associated with the stability of the Leggett and Higgs modes, because the resonance widths are associated with lifetimes of excitations. 
To explore this, here we have numerically
solved the gap Eqs.(\ref{self_consistency}, \ref{gap_eqn}) 
incorporating the effect of interband coupling $\lambda_I$ to
 obtain $L(\omega)$ and $H_\gamma(\omega)$ at finite temperatures 
with Eq.(\ref{function_F}), for the values of parameters 
estimated for $\mathrm{MgB_2}$, $\lambda_I=-0.19$, etc [see section \ref{Leggett_mode}].  
In Fig. \ref{temp_dep}, panel (a) depicts the temperature dependence of the gap 
energies against temperature. We then plot $|L(\omega)|$ and $|H_\gamma(\omega)|$ at several 
temperatures as indicated by horizontal lines in panel (a).  In 
panel (b) we can see that the peak of $|L(\omega)|$ becomes sharper 
as temperature increases, which indicates a stabilization of the Leggett mode. This contrasts 
with the peak of $|H_\alpha(\omega)|$ at $\omega=2\Delta_\alpha$ in panel (c), 
which is weakened and finally disappears with increasing temperature.  
The peak of $|H_\beta(\omega)|$ at $\omega=2\Delta_\beta$ in (d) 
remains sharp even at finite temperatures.  

\begin{figure}
	\begin{center}
		\includegraphics[width=8cm,clip]{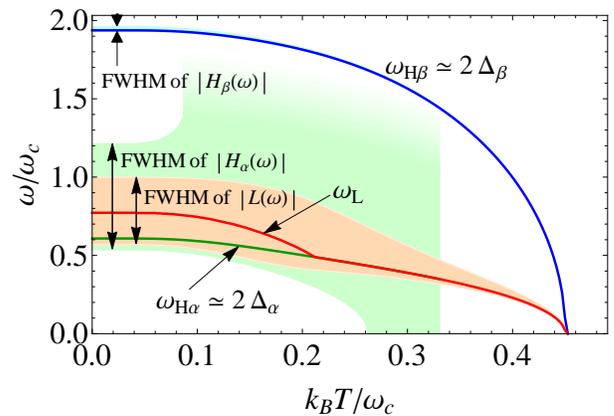}

		\caption{Temperature dependence of the Leggett mode energy $\omega_{\mathrm{L}}$ (red curve) and the Higgs mode energies $\omega_{\mathrm{H}\alpha}$ (green), and $\omega_{\mathrm{H}\beta}$ (blue). Their resonance widths are 
indicated by orange, green and blue shadings, respectively. The green line denoting $\omega_{\mathrm{H}\alpha}$ overlaps with the red line up to 
the right edge of the green region, at which the green line is terminated. Used parameters are the same as Fig. \ref{temp_dep}.}\label{temp_dep2}
	\end{center}
\end{figure}

We summarize these by plotting the 
peak positions and widths for $|L(\omega)|$ and $|H_\gamma(\omega)|$ against temperature in Fig. \ref{temp_dep2}.
At $T=0$, the Leggett mode has an energy $\omega_{\mathrm{L}}$ between the two superconducting gaps $2\Delta_\alpha$ and $2\Delta_\beta$, and has a broad width. As $\omega_{\mathrm{L}}$ decreases with increasing temperature, the mode energy reaches the lower gap $2\Delta_\alpha$ at a certain point. At even higher temperatures, 
the mode energy $\omega_{\mathrm{L}}$ traces $2\Delta_\alpha$ with a slightly narrowing width. 
As for the Higgs modes, their energies follow the temperature dependence of the gaps, 
which can be understood in terms of Eq.(\ref{H_factor}) with $G_\gamma(\omega=2\Delta_\gamma)=0$. The Higgs mode with higher energy has quite narrow widths for the whole temperature range,
which reveals that the Higgs mode with the higher energy remains long-lived.
On the other hand, the width of the Higgs mode with lower energy is broadened as temperature is increased, and the peak of this mode disappears at a certain temperature.

Sharpening of the Leggett mode might seem to arise because the mode becomes one of the lowest-energy excitations and thus a stable mode at high temperatures. However, this cannot explain the broadening and disappearance of the lower-energy Higgs mode, which is also degenerate with the lowest-energy excitations. Rather, this can be intuitively explained as follows. 
At low temperatures, Cooper pairs 
are basically formed through the intraband interactions, giving rise to Higgs modes primarily associated with each band (although modifications due to the interband coupling, such as a broadening of the lower-energy Higgs mode, exist).  
At higher temperatures, however, Cooper pairs with the smaller gap could no longer be formed if it were not for the interband coupling, since the single-band BCS critical temperature is lower for the smaller gap. Through the interband coupling, the larger gap $2\Delta_\beta$ makes the lower $2\Delta_\alpha$ finite even at higher temperatures, but Cooper pairs lack $\alpha$-band character there, and only the Higgs mode associated with the larger gap remains.

\section{Third-harmonic generation from light-induced collective modes} \label{sec_THG}

Matsunaga \textit{et al.}\cite{Matsunaga2} have experimentally revealed that a conventional superconductor NbN illuminated with an intense THz wave emits the third harmonics.
This is an intrinsic nonlinear phenomenon in superconductors\cite{Tsuji, Cea}.
As mentioned in Sec.\ref{section2}, Anderson's pseudospins respond to $A(t)^2$.
Such pseudospin motions induce an electric current which is itself proportional to $A(t)$ (as shown later). 
The induced current thus follows $\sim A(t)^3$ in total, and the third-harmonic generation (THG) emerges, which has been detected by a simple transmission experiment.
In single-band superconductors, THG is resonantly enhanced when the doubled frequency of the pump light coincides with the superconducting gap $2\Delta$, 
where the THG resonance occurs due to both the Higgs mode and density fluctuations, the latter being shown to be dominant within the BCS theory\cite{Cea}.  
We can then raise an intriguing question of how THG should 
look like in two-band superconductors. 

Amplitude of the emitted electric field is proportional to the induced current, so that we can concentrate on the light-induced third-order current, 
which is, for the two-band case,  expressed as
\begin{align}
\bm{j}&=e\sum_{\gamma\bm{k}\sigma}\left[\nabla_{\bm{k}}\epsilon_{\gamma(\bm{k}-e\bm{A})}\right]\gamma^\dag_{\bm{k}\sigma}\gamma_{\bm{k}\sigma} \notag \\
&=e\sum_{\gamma\bm{k}}\left[\nabla_{\bm{k}}\epsilon_{\gamma(\bm{k}-e\bm{A})}-\nabla_{\bm{k}}\epsilon_{\gamma(\bm{k}+e\bm{A})}\right]\left(\sigma^z_{\gamma\bm{k}}+\frac{1}{2}\right) \notag \\
&\quad+e\sum_{\gamma\bm{k}}\left[\nabla_{\bm{k}}\epsilon_{\gamma(\bm{k}-e\bm{A})}+\nabla_{\bm{k}}\epsilon_{\gamma(\bm{k}+e\bm{A})}\right]\left(\sigma^e_{\gamma\bm{k}}-\frac{1}{2}\right), \label{current}
\end{align}
where 
\begin{equation}
\sigma^e_{\gamma\bm{k}}=\frac{1}{2}
(\begin{array}{cc}
\gamma^\dag_{\bm{k}\uparrow} & \gamma_{-\bm{k}\downarrow}
\end{array})\bm{1}\left(
\begin{array}{c}
\gamma_{\bm{k}\uparrow} \\ \gamma^\dag_{-\bm{k}\downarrow}
\end{array}\right)
\end{equation}
remains constant during time evolution in the mean-field Hamiltonian (\ref{Ham_sum}).
Because $\bm{\sigma}_{\gamma\bm{k}}(t)$ responds to $A(t)^2$, forced 
oscillations of $\delta\sigma^z_{\gamma\bm{k}}(t)$ have a frequency $2\Omega$ for the incident frequency $\Omega$ of $A(t)$.
The first term in Eq.(\ref{current}) thus accommodates a third-harmonic component with a frequency $3\Omega$, 
\begin{equation}
\bm{j}^{(3)}(t)=-2e^2A(t)\sum_\gamma\sum_{\bm{k}}\frac{\partial^2\epsilon_{\gamma\bm{k}}}{\partial \bm{k}\partial k_x}\delta\sigma^z_{\gamma\bm{k}}(t).
\end{equation}
According to Eq.(\ref{linearized}),
\begin{align}
&\bm{j}^{(3)}(t)=-2e^2A(t)\int \frac{d\omega}{2\pi}e^{i\omega t}\sum_\gamma\sum_{\bm{k}}\frac{\partial^2\epsilon_{\gamma\bm{k}}}{\partial \bm{k}\partial k_x}\frac{\sigma^x_{\gamma\bm{k}}}{4E_{\gamma\bm{k}}^2-\omega^2} \notag \\
&\times\left[4\epsilon_{\gamma\bm{k}}\delta\Delta'_\gamma(\omega)+2i\omega\delta\Delta''_\gamma(\omega)
-4\Delta_\gamma\delta b^z_{\gamma\bm{k}}(\omega)\right].
\label{3rd_current}
\end{align}
The first term in the bracket on the right-hand side corresponds to the Higgs-mode contribution\cite{Tsuji}, while the last term the contribution from density fluctuations\cite{Cea}. The second term gives a phase contribution, 
which is related to the Leggett mode as discussed later. 
We denote the contributions from the Higgs mode, phase (Leggett mode) and density fluctuations as $\bm{j}^{(3)}_{\mathrm{H}}$, $\bm{j}^{(3)}_{\mathrm{L}}$ and $\bm{j}^{(3)}_{\mathrm{d}}$, respectively, with 
the total third-order current 
\begin{equation}
\bm{j}^{(3)}(t)=\bm{j}^{(3)}_{\mathrm{H}}(t)+\bm{j}^{(3)}_{\mathrm{L}}(t)+\bm{j}^{(3)}_{\mathrm{d}}(t).
\end{equation}

From Eq.(\ref{H_solution}) and a function (\ref{function_X}), the Higgs-mode contribution is reduced to
\begin{align}
j^{(3)i}_{\mathrm{H}}(t)&=-e^4A(t)\sum_{\gamma}\frac{c^i_{\gamma1}}{c_{\gamma1}}D_\gamma \notag \\
&\quad\times\int\frac{d\omega}{2\pi}e^{i\omega t}A^2(\omega)H_\gamma(\omega)X_\gamma(\omega),
\label{current_H}
\end{align}
where $c^i_{\gamma1}$ is defined by
\begin{equation}
\sum_{\bm{k}}\delta(\epsilon-\epsilon_{\gamma\bm{k}})\frac{\partial^2\epsilon_{\gamma\bm{k}}}{\partial k_i\partial k_x}
=D_\gamma(c^i_{\gamma0}+c^i_{\gamma1}\epsilon+\cdots).
\label{expansion2}
\end{equation}
As before, $x$ is the polarization direction of the incident light.
For $i=x$, Eq.(\ref{expansion2}) is equivalent to Eq.(\ref{expansion}), hence $c^x_{\gamma n}=c_{\gamma n}$ $(n=0,1,\cdots)$.
While $X_\gamma(\omega)$ has no distinctive spectral feature, $H_\gamma(\omega)$ has peak structures at $\omega=2|\Delta_{\alpha, \beta}|$ which represent the two Higgs modes, so that the THG arising from Eq.(\ref{current_H}) is resonantly enhanced at $2\Omega\approx2|\Delta_{\alpha, \beta}|$.
For strong interband interactions for which the Higgs mode with the lower energy is broadened,
the corresponding THG will be weaker than the higher-energy one.

Now we turn to the other contributions.
Using Eqs.(\ref{driving_force}, \ref{solution}, \ref{function_Y}, \ref{L_factor}),
the phase and density-fluctuation contributions are reduced, respectively, to
\begin{align}
&j^{(3)i}_{\mathrm{L}}(t)=4e^4A(t) \nonumber \\
&\quad\times
\left[(c_{\alpha0}^i-c_{\beta0}^i)\frac{\lambda_I\sqrt{D_\alpha D_\beta}\Delta_\alpha\Delta_\beta}{\det\lambda}
\int\frac{d\omega}{2\pi}e^{i\omega t}A^2(\omega)L(\omega)\right. \nonumber \\
&\quad\quad\left.-\sum_\gamma c_{\gamma0}^ic_{\gamma0}D_\gamma\Delta_\gamma
\int\frac{d\omega}{2\pi}e^{i\omega t}A^2(\omega)F_\gamma(\omega)\right], \label{2nd} \\
&j^{(3)i}_{\mathrm{d}}(t)=4e^4A(t)\sum_\gamma\tilde{c}_{\gamma0}^iD_\gamma\Delta_\gamma
\int\frac{d\omega}{2\pi}e^{i\omega t}A^2(\omega)F_\gamma(\omega), \label{3rd}
\end{align}
with $c_{\gamma0}^i$ defined by Eq.(\ref{expansion2}) and $\tilde{c}_{\gamma0}^i$ by
\begin{equation}
\sum_{\bm{k}}\delta(\epsilon-\epsilon_{\gamma\bm{k}})\frac{\partial^2\epsilon_{\gamma\bm{k}}}{\partial k_i\partial k_x}\frac{\partial^2\epsilon_{\gamma\bm{k}}}{\partial k_x^2}
=D_\gamma(\tilde{c}^i_{\gamma0}+\tilde{c}^i_{\gamma1}\epsilon+\cdots).
\label{expansion3}
\end{equation}
We can show that, when we take account of the screening due to the long-range Coulomb interaction, the term represented by the third line in Eq.(\ref{2nd}) is screened out, 
while the same term appears as a screening effect 
in Eq.(\ref{3rd}).  
Thus we have a screened form as
\begin{align}
j^{(3)i}_{\mathrm{L}}(t)&=4e^4A(t)(c_{\alpha0}^i-c_{\beta0}^i)\frac{\lambda_I\sqrt{D_\alpha D_\beta}\Delta_\alpha\Delta_\beta}{\det\lambda}
\notag \\
&\quad\times\int\frac{d\omega}{2\pi}e^{i\omega t}A^2(\omega)L(\omega), \label{current_L} \\
j^{(3)i}_{\mathrm{d}}(t)&=4e^4A(t)\sum_\gamma\left(\tilde{c}_{\gamma0}^i-c_{\gamma0}^ic_{\gamma0}\right)
D_\gamma\Delta_\gamma
\notag \\
&\quad\times\int\frac{d\omega}{2\pi}e^{i\omega t}A^2(\omega)F_\gamma(\omega), \label{current_d}
\end{align}
while the Higgs-mode contribution (\ref{current_H}) is not modified by screening.

Equation (\ref{current_L}) can be reduced to
\begin{align}
j^{(3)i}_{\mathrm{L}}(t)&=-4e^2A(t)(c_{\alpha0}^i-c_{\beta0}^i)\frac{\lambda_I\sqrt{D_\alpha D_\beta}\Delta_\alpha\Delta_\beta}{\det\lambda}
\notag \\
&\quad\times\int^t_{-\infty}dt'\delta\left[\theta_\alpha(t')-\theta_\beta(t')\right]
\label{current_L3}
\end{align}
with the use of Eq.(\ref{Leggett_solution}). As already shown, forced oscillation of the phase difference $\delta\left[\theta_\alpha(t)-\theta_\beta(t)\right]$ resonates with the Leggett mode at $2\Omega\approx\omega_{\mathrm{L}}$, so that Eq.(\ref{current_L3}) and the resulting third-harmonic is enhanced on the same condition. 
Thus this current represents {\it THG from resonantly excited Leggett mode}. Presence of this phase contribution sharply contrasts with the single-band cases where the phase contribution is fully screened out.

On the other hand, Eq.(\ref{current_d}) for $\bm{j}^{(3)}_{\mathrm{d}}$ is resonantly enhanced at $2\Omega\approx2|\Delta_{\alpha,\beta}|$, 
because $F_\gamma(\omega)$ diverges at $\omega=2|\Delta_\gamma|$ following $F_\gamma(\omega)\sim1/\sqrt{4\Delta_\gamma^2-\omega^2}$ [which can be seen in Eq.(\ref{simplest_F})].  
Following Ref.\onlinecite{Cea}, we call this a 
``density fluctuation"
(because $F_\gamma(\omega)$ has a form of density-density correlation function).

Now, let us display a model calculation for THG.
We adopt the two-dimensional square lattice with a band dispersion
$\epsilon_{\gamma\bm{k}}=\epsilon_\gamma^0-2t_\gamma(\cos k_\xi+\cos k_\eta)$ 
with an offset $\epsilon_\gamma^0$ for each band, 
where $\xi$ and $\eta$ denote crystal axes, which are in general 
not parallel to $x$ (the polarization of laser).
We take 
$\epsilon_\alpha^0=2.5$, $t_\alpha=1$, $\epsilon_\beta^0=-4$, $t_\beta=1.5$, 
which correspond to 
$D_\alpha=0.099$, $D_\beta=0.064$, 
$c_{\alpha0}=1.25$, $c_{\beta0}=-2$, 
$c_{\alpha1}=-0.28$, $c_{\beta1}=-0.28$  
(with analytic expressions for these parameters given in Appendix B),
to make the ratio between $D_\alpha$ and $D_\beta$ close to that of MgB$_2$.

When the incident wave is polarized along the $[1,0]$ direction (i.e. $\xi=x$, $\eta=y$), 
we obtain $\tilde{c}^x_{\alpha0}=1.8$, $\tilde{c}^x_{\beta0}=4.5$.
On the other hand, 
when the incident wave is polarized along $[1,1]$, we have 
exactly $\tilde{c}^x_{\gamma0}=c_{\gamma0}^2$, 
so that Eq.(\ref{current_d}) vanishes for $i=x$.
In this case, we must take account of $\tilde{c}^x_{\gamma2}$ and $c_{\gamma2}$ 
to precisely calculate the density-fluctuation contribution:
\begin{align}
&j^{(3)x}_{\mathrm{d}}(t)
=-4e^4A(t)\sum_\gamma\left(\frac{\tilde{c}^x_{\gamma2}}{4}-\frac{c_{\gamma0}c_{\gamma2}}{2}\right)
D_\gamma\Delta_\gamma \nonumber\\
&\times\int\frac{d\omega}{2\pi}e^{i\omega t}A^2(\omega)
\left[\frac{\lambda_{\overline{\gamma}\overline{\gamma}}\Delta_\gamma
-\lambda_{\gamma\overline{\gamma}}\Delta_{\overline{\gamma}}}
{\det\lambda}+G_\gamma(\omega)\right].
\label{subleading}
\end{align}
Numerical calculation gives
$\tilde{c}^x_{\alpha2}/4-c_{\alpha0}c_{\alpha2}/2\simeq\tilde{c}^x_{\beta2}/4-c_{\beta0}c_{\beta2}/2\simeq0.046$
for the $[1,1]$ direction.

For simplicity, we consider monochromatic illumination, $A(-\infty<t<\infty)=A_0\sin\Omega t$, i.e.,
$A^2(\omega)=\frac{\pi}{2}A_0^2\left[2\delta(\omega)-\delta(\omega-2\Omega)-\delta(\omega+2\Omega)\right]$.
For the pairing interaction, we take the values for MgB$_2$, i.e. 
$\lambda_{\alpha\alpha}=-0.28$, $\lambda_{\beta\beta}=-0.96$, $\lambda_I=-0.19$.

\begin{figure}
	\begin{center}
		\includegraphics[width=7.5cm,clip]{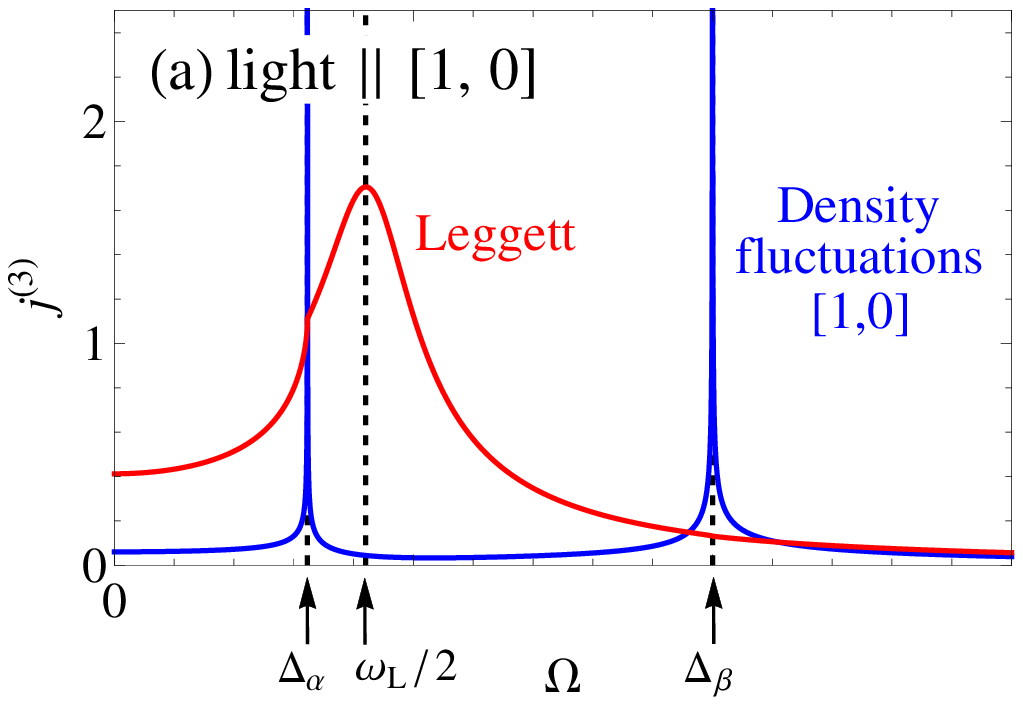}
		\includegraphics[width=7.55cm,clip]{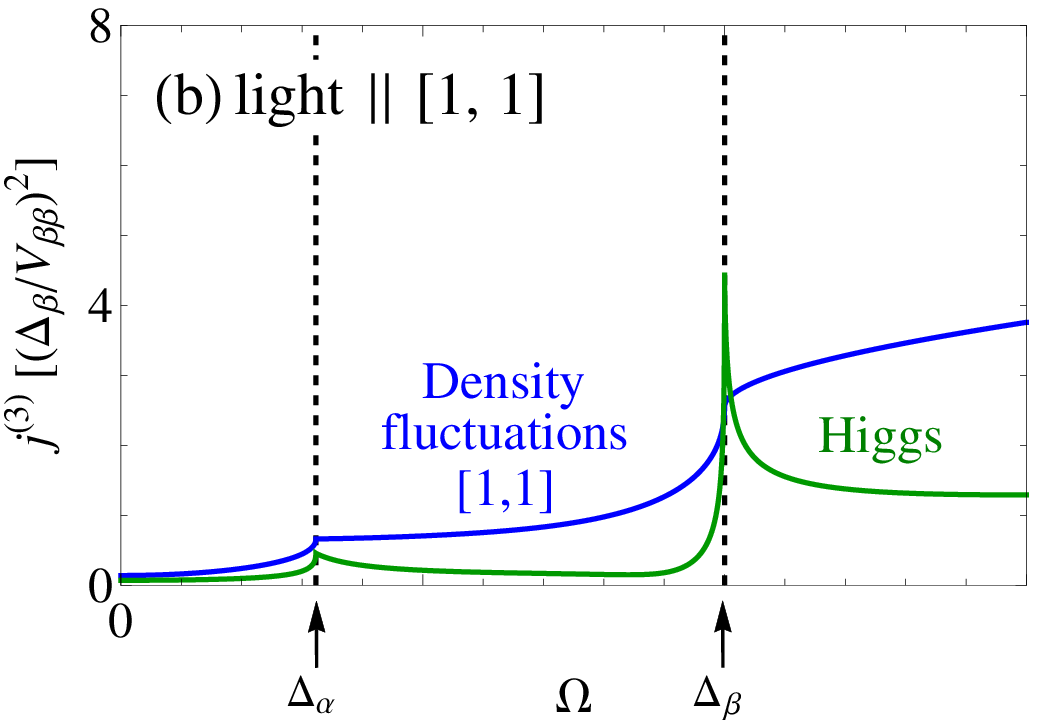}

		\caption{Amplitude of the THG spectra for the incident light polarized along the
$[1,0]$ (a)
and $[1,1]$ (b) directions of the square lattice.
In (a), the red and blue curves respectively show $|j^{(3)x}_{\mathrm{L}}|$ and
$|j^{(3)x}_{\mathrm{d}}|$,
while in (b), the blue and green curves respectively show $|j^{(3)x}_{\mathrm{d}}|$ and
$|j^{(3)x}_{\mathrm{H}}|$.  Note that 
$|j^{(3)x}_{\mathrm{L}}|$ and $|j^{(3)x}_{\mathrm{H}}|$ have no 
polarization dependence.
The vertical axis in (b) is normalized by 
$(\Delta_\beta/V_{\beta\beta})^2$.
The positions of $\Omega=\Delta_\alpha,~\omega_{\mathrm{L}}/2,~\Delta_\beta$
are respectively indicated by dashed lines and arrows.
The parameters used here are explained in the text.
}
\label{THG}
	\end{center}
\end{figure}

In Fig. \ref{THG}(a), we show $j^{(3)x}_{\mathrm{L}}$ and
$j^{(3)x}_{\mathrm{d}}$ for the case of the incident light
polarized along the $[1,0]$ direction.
In this case, the Higgs mode contribution is, within the 
present mean-field formalism, subleading by a factor
$(\Delta/V)^2$,
where $\Delta$ and $V$ are the energy scales of the superconducting gap and
the paring interaction, respectively.
This is consistent with the analysis for the single-band case \cite{Cea}.
One can see a prominent Leggett-mode contribution with a relatively 
broad resonance peak of THG at $\Omega=\omega_L/2$, 
besides sharp peaks at $\Omega=\Delta_\alpha, \Delta_\beta$ due mainly to the
density fluctuations.
The Leggett- and Higgs-mode contributions do not have polarization
dependence, while
the density fluctuation sensitively depends on the polarization.
When the incident light is polarized along the $[1,1]$ direction,
the magnitude of the density-fluctuation reduces by a factor
$(\Delta/V)^2$,
and becomes 
comparable with that of the Higgs mode [Fig. \ref{THG}(b)]. 
When $\Delta\sim$ meV and $V\sim$ eV as in MgB$_2$, 
this factor should become extremely small in the present 
mean-field treatment, so that the leading contribution comes 
from the Leggett mode. The resonance peaks at $\Omega=\Delta_\alpha,
\Delta_\beta$ will be mainly generated by the Higgs mode, 
because the density-fluctuation contribution does not show resonance
for this polarization direction [Fig. \ref{THG}(b)].

In general, relative importance of the Leggett mode and density fluctuations in THG should depend on the band structure.
In Fig. \ref{THG}(a), density fluctuations are seen to be much screened, 
because the band structure adopted there 
is nearly parabolic around the Fermi energy 
(i.e., for an isotropic parabolic band, 
the coefficient $\tilde{c}_{\gamma0}-c_{\gamma0}^2$ in Eq.(\ref{current_d}) vanishes)\cite{Cea2}.
In addition, the model adoped here has an electron-like band ($\alpha$) and a hole-like band ($\beta$),
so that the coefficient $c_{\alpha0}-c_{\beta0}$ in the Leggett-mode contribution (\ref{current_L}) is large 
(recall that $c_{\gamma0}$ is approximately equal to the inversed mass).
These are the reasons why the Leggett-mode contribution is prominent 
in Fig. \ref{THG}(a), 
with the situation being 
similar to Raman scattering in MgB$_2$ \cite{Blumberg}.

As for the Higgs modes,
effects beyond the mean-field theory (such as retardation and
correlation effects) have recently been proposed to enhance the relative
importance of their
contribution to THG\cite{Tsuji2}. Specifically, the Higgs-mode contribution
is shown, for the single-band case with
general polarization, to be not necessarily subject
to the reduction by the factor of $(\Delta/V)^2$.
It would be interesting if similar effects
arise in multiband systems as well.
When the Higgs-mode contributions are observable,
they should exhibit
the lower-energy resonance at $\Omega=\Delta_\alpha$ broader and weaker
than the higher-energy one
at $\Omega=\Delta_\beta$
as described above, which
contrasts with the density-fluctuation contribution
with both peaks sharp.
Therefore, a line-shape analysis of THG resonance may help one to resolve 
the origin of observed data.

Another possibility for experimentally 
distinguishing the collective modes from density fluctuations is 
the dependence of THG on the direction 
of the polarization of light\cite{Cea}.
As seen in Fig. \ref{THG}, 
the density-fluctuation contribution strongly depends on the polarization direction of the incident light, 
while the Higgs- and Leggett-mode contributions do not, 
as exemplified here for the square lattice. Polarization dependence in real materials should be dominated by actual band structures,
but, roughly speaking, one can expect smaller polarization dependence of THG 
for collective modes than for density fluctuations, 
because collective modes are basically isotropic excitations (in $s$-wave superconductors) 
while density fluctuations are generally not.

\section{Conclusion}
We have investigated collective modes resonantly excited by electromagnetic waves for two-band superconductors having different BCS gap 
energies. For weaker interband pairing interactions, there emerge three collective modes that can be optically excited: two Higgs modes corresponding  to amplitude oscillation of two order parameters, and the Leggett mode corresponding to oscillation of the relative phase. For stronger interband interactions, which should include the case of $\mathrm{MgB_2}$, the Leggett mode and one of the Higgs modes are destabilized with their resonances weakened. At finite temperatures, the Leggett mode slightly recovers its stability, while the Higgs mode associated with the smaller gap disappears; the Higgs mode with the larger energy always remains long-lived.  
We further find that all of these collective modes contribute to
the third-harmonic generation (THG).
Specifically, we have shown that the Leggett mode can be observable in THG experiments.
Density fluctuations also contribute to THG and have the same resonance frequency as the Higgs modes.
A difference is that THG from the lower-energy Higgs mode is weakened by the interband interaction, 
while the lower-energy peak from density fluctuations is not.
Such a difference along with the polarization dependence may help one to experimentally distinguish contributions from the Higgs modes and density fluctuations.

Processes beyond the mean-field approximation, such as retardation and 
correlation effects, may significantly affect the mode properties and nonlinear response\cite{Tsuji2}.
An even more interesting possibility is the interaction between the Leggett and the Higgs modes\cite{Krull}. 
When we consider higher-order processes beyond the linearized equation of motion, a coupling between these collective modes should appear, which is expected to lead to further features in the dynamical behavior of the order parameters. 
While we have concentrated on the linear regime here, the nonlinear couplings between coexisting collective modes will serve as an intriguing future problem.

Detailed experiments on the Leggett and Higgs modes will 
be desirable. 
In the case of MgB$_2$, a terahertz wave will be suited for resonant optical excitations, 
because the superconducting gaps in this material lie in the milielectronvolt energy scale. Probing the collective 
modes may also be possibly applicable to the iron pnictides which are also
multiband with a similar energy scale.
In a broader context, a message is that the analysis of collective modes 
is expected to pave a new pathway for probing the condensates in 
multiband superconductors with intraband and interband pairing interactions, and a possibility of controlling superconductivity.

\begin{acknowledgments}
We wish to thank A. J. Leggett for a valuable comment.  
We are also benefitted from illuminating discussions with 
R. Shimano, R. Matsunaga and K. Tomita.  
Discussions with K. Kuroki, M. Yamada, and 
A. Sugioka are also gratefully acknowledged.  
The present work was supported by  JSPS KAKENHI Grant JP26247057 and ImPACT project (No. 2015-PM12-05-01) from the Cabinet Office, Japan.
N.T. is supported by JSPS KAKENHI Grants JP25104709 and JP25800192.\\
\end{acknowledgments}

\appendix
\section{Table of symbols}

Let us first summarize the symbols used in the paper in Table \ref{symbols}.

\begin{table}[h]
\caption{Summary of symbols.} \label{symbols}
\begin{tabular}[b]{cl} \hline \hline
Symbol & Definition \\ \hline \hline
$\gamma=\alpha$ or $\beta$ & Band indices; annihilation operators \\ \hline
$\Delta_\gamma$ & Superconducting gaps [Eq.(\ref{gap})] \\ \hline
$\lambda_{\gamma\gamma'}$ & Dimensionless paring interaction [Eq.(\ref{dlpi})] \\ \hline
$\lambda_I$ & Dimensionless interband coupling [Eq.(\ref{dlipi})] \\ \hline
$\det\lambda$ & $\lambda_{\alpha\alpha}\lambda_{\beta\beta}-\lambda_{\alpha\beta}\lambda_{\beta\alpha}$ [Eq.(\ref{dlpi})] \\ \hline
$\bm{\sigma}_{\gamma\bm{k}}$ & Anderson's pseudospin [Eq.(\ref{pseudospin})] \\ \hline
$\Omega$ & Frequency of incident light \\ \hline
\multirow{2}{*}{$\omega_{\mathrm{H}\gamma}$} & Energy of Higgs modes, i.e., peak position of\\
& $\;\;|H_\gamma(\omega)|$, equal to $2\Delta_\gamma$ [Sec. \ref{Higgs_modes}] \\ \hline
\multirow{2}{*}{$\omega_{\mathrm{L}}$} & Energy of Leggett mode, i.e.,\\
& $\;\;$ peak position of $|L(\omega)|$ [Sec. \ref{Leggett_mode}] \\ \hline
\multirow{2}{*}{$A^2(\omega)$} & Fourier transform of \\
& $\;\;$ squared vector potential $A(t)^2$ \\ \hline
$\bm{b}_{\gamma\bm{k}}$ & Pseudomagnetic field [Eq.(\ref{pseudomagnetic})] \\ \hline
\multirow{2}{*}{$c_{\gamma0},~c_{\gamma1}$} & Expansion coefficients of \\
& $\;\;\sum_{\bm{k}}\delta(\epsilon-\epsilon_{\gamma\bm{k}})\frac{\partial^2\epsilon_{\gamma\bm{k}}}{\partial k_x^2}$ [Eq.(\ref{expansion})] \\ \hline
\multirow{2}{*}{$c^i_{\gamma0}$} & Expansion coefficient of \\
& $\;\;\sum_{\bm{k}}\delta(\epsilon-\epsilon_{\gamma\bm{k}})\frac{\partial^2\epsilon_{\gamma\bm{k}}}{\partial k_i\partial k_x}$ [Eq.(\ref{expansion2})] \\ \hline 
\multirow{2}{*}{$\tilde{c}^i_{\gamma0}$} & Expansion coefficient of \\
& $\;\;\sum_{\bm{k}}\delta(\epsilon-\epsilon_{\gamma\bm{k}})\frac{\partial^2\epsilon_{\gamma\bm{k}}}{\partial k_i\partial k_x}\frac{\partial^2\epsilon_{\gamma\bm{k}}}{\partial k_x^2}$ [Eq.(\ref{expansion3})] \\ \hline 
$D_\gamma$ & Density of states on Fermi surface [Eq.(\ref{DOS})] \\ \hline
$E_{\gamma\bm{k}}$ & Bogoliubov quasi-particle's energy [Eq.(\ref{QPE})]\\ \hline
\multirow{2}{*}{$F_\gamma(\omega)$} & Eq.(\ref{function_F2}) or (\ref{function_F}): \\
& $\;\;$ resonance factor for density fluctuations \\ \hline 
$G_\gamma(\omega)$ & Eq.(\ref{function_G}): $(4\Delta_\gamma^2-\omega^2)F_\gamma(\omega)$ \\ \hline
$H_\gamma(\omega)$ & Eq.(\ref{H_factor}): resonance factor for Higgs modes \\ \hline
$L(\omega)$ & Eq.(\ref{L_factor}): resonance factor for Leggett mode \\ \hline
$V_{\gamma\gamma'}$ & Paring interaction [Eq.(\ref{H_2band})] \\ \hline
$\det V$ & $V_{\alpha\alpha}V_{\beta\beta}-V_{\alpha\beta}V_{\beta\alpha}$ \\ \hline
\multirow{2}{*}{$X_\gamma(\omega)$} & Eq.(\ref{function_X}): for nonlinear coupling \\
& $\;\;$ between light and Higgs modes \\ \hline
\multirow{2}{*}{$Y_\gamma(\omega)$} & Eq.(\ref{function_Y2}) or (\ref{function_Y}): for nonlinear coupling\\
& $\;\;$ between light and Leggett mode \\ \hline \hline
\end{tabular}
\end{table}

\section{Band parameters}

Here we give analytic expressions for the parameters in terms 
of the model band dispersion
used in Sec. \ref{sec_THG}.
We have adopted the square lattice with a two-dimensional band dispersion, 
\begin{equation}
\epsilon_{\bm{k}}=\epsilon_0-2t(\cos k_x+\cos k_y),\label{model}
\end{equation}
where the band index $\gamma$ is omitted for simplicity.
Density of states, defined by Eq.(\ref{DOS}), then reduces to
\begin{equation}
D(\epsilon)=\frac{1}{2\pi^2t}K\left(1-\frac{(\epsilon-\epsilon_0)^2}{16t^2}\right),
\end{equation}
where
\begin{equation}
K(m)=\int_0^{\pi/2}d\theta\frac{1}{\sqrt{1-m\sin^2\theta}}
\end{equation}
is the complete elliptic integral of the first kind.
The value on the Fermi surface is thus given as
\begin{equation}
D=\frac{1}{2\pi^2t}K\left(1-\frac{\epsilon_0^2}{16t^2}\right).\label{D}
\end{equation}
The coefficients $c_0,c_1$ defined by Eq.(\ref{expansion}) become
\begin{align}
c_0&=\frac{\epsilon_0}{2},\label{c0}\\
c_1&=-\frac{1}{2}\left[1-\frac{\epsilon_0}{D}\frac{dD(0)}{d\epsilon}\right] \nonumber\\
&=-\frac{1}{2\left(1-\frac{\epsilon_0^2}{16t^2}\right)}
\left[1-\frac{E\left(1-\frac{\epsilon_0^2}{16t^2}\right)}{K\left(1-\frac{\epsilon_0^2}{16t^2}\right)}\right],\label{c1}
\end{align}
where
\begin{equation}
E(m)=\int_0^{\pi/2}d\theta\sqrt{1-m\sin^2\theta}
\end{equation}
is the complete elliptic integral of the second kind, and
\begin{equation}
\frac{dK(m)}{dm}=\frac{1}{2m(1-m)}\left[E(m)-(1-m)K(m)\right]
\end{equation}
has been used.
Finally, $\tilde{c}_0$ defined by Eq.(\ref{expansion3}) is given as
\begin{equation}
\tilde{c}_0=8t^2
\left[\frac{1}{2}+\frac{\epsilon_0^2}{16t^2}
-\frac{E\left(1-\frac{\epsilon_0^2}{16t^2}\right)}{K\left(1-\frac{\epsilon_0^2}{16t^2}\right)}\right].\label{ctilde0}
\end{equation}
Equations (\ref{D}), (\ref{c0}), (\ref{c1}), (\ref{ctilde0}) give 
analytic expressions for the required parameters.

When the polarization direction of the incident light is rotated from the crystal axes by an angle $\theta$, 
electron momentum in Eq.(\ref{model}) is replaced by
\begin{equation}
k_x\to k_x\cos\theta-k_y\sin\theta,~
k_y\to k_x\sin\theta+k_y\cos\theta.
\end{equation}
Even in this case, $c_0$ and $c_1$ do not depend on the direction of polarization,
and thus are given by Eqs.(\ref{c0}) and (\ref{c1}), respectively.
By contrast, $\tilde{c}_0-c_0^2$ is modified as
\begin{equation}
\tilde{c}_0-c_0^2=8t^2
\left[\frac{1}{2}+\frac{\epsilon_0^2}{32t^2}
-\frac{E\left(1-\frac{\epsilon_0^2}{16t^2}\right)}{K\left(1-\frac{\epsilon_0^2}{16t^2}\right)}\right]\cos^22\theta,
\end{equation}
and therefore vanishes for $\theta=\pi/4$. 
In that case, values of $c_2$ and $\tilde{c}_2$
are necessary to evaluate the density-fluctuation contribution, which is given by Eq.(\ref{subleading}).
Because the analytic expressions for $c_2$ and $\tilde{c}_2$ are complicated, 
we have calculated them numerically.

\end{document}